\documentclass[a4paper,12pt]{article}

\usepackage[a4paper,
  left=2.5cm, right=2.5cm,
  top= 3cm, bottom=4cm]{geometry}
\usepackage{amsmath}
\usepackage{oldgerm}
\usepackage{amssymb}
\usepackage{bbm}
\usepackage[dvips]{graphicx}
\usepackage{epsfig}
\usepackage{color}
\usepackage{cite}
\usepackage{epic}
\usepackage{hyperref}

\usepackage[T1]{fontenc}
\usepackage[english]{babel}
\numberwithin{equation}{section}

\newcommand{\be}{\begin{equation}}  
\newcommand{\ee}{\end{equation}}

\newcommand{\rem}[1]{} 

\def\deg{\text{deg }}
\def\C{\mathbb{C}}
\def\Z{\mathbb{Z}}

\def\P{\mathbb{P}}

\def\id{\operatorname{id}}

\def\Hirz[#1]{\mathbbm{F}_{#1}}
\def\o[#1]{\overline{#1}}

\newcommand{\red}[1]{}
\renewcommand{\red}[1]{{\color{red}  {#1}}}

\def\a{\mathbf{a}}

\begin{document}

\begin{titlepage}

\vskip -0.5cm

\begin{flushright}

\end{flushright}
 
\vskip 1cm
\begin{center}
 
{\large \bf $G_4$ Flux, Algebraic Cycles and  \\  \vspace{2mm} Complex Structure Moduli Stabilization} 
 
 \vskip 1.2cm
 
 A.~P.~Braun$^a$  and R.~Valandro$^b$

 \vskip 0.4cm
 
{\it $^a$Department of Mathematical Sciences, Durham University, Lower Mountjoy, 
Stockton Rd, Durham DH1 3LE, UK \\[2mm]
 
 $^b$Dipartimento di Fisica, Universit\`a di Trieste, Strada Costiera 11, I-34151 Trieste, Italy \\
and INFN, Sezione di Trieste, Via Valerio 2, I-34127 Trieste, Italy	
 }
 \vskip 1.5cm
 
\abstract{We construct $G_4$ fluxes that stabilize all of the 426 complex 
structure moduli of the sextic Calabi-Yau fourfold at the Fermat point. Studying 
flux stabilization usually requires solving Picard-Fuchs equations, which 
becomes unfeasible for models with many moduli. Here, we 
instead start by considering a specific point in the complex structure moduli 
space, and look for a flux that fixes us there. We show how to 
construct such fluxes by using algebraic cycles and analyze flat 
directions. This is discussed in detail for the sextic Calabi-Yau fourfold at 
the Fermat point, and we observe that there appears to be tension between 
M2-tadpole cancellation and the requirement of stabilizing all moduli.
Finally, we apply our results to show that even though symmetric fluxes allow 
to automatically solve most of the F-term equations, they typically lead 
to flat directions.
} 

\end{center}

\end{titlepage}

\tableofcontents

\newpage

\section{Introduction}

M-theory compactified on a Calabi-Yau (CY) fourfold $X$ has $h^{1,3}(X)$ complex 
structure moduli, which can be thought of as variations of the holomorphic top 
form $\Omega$. In such models, one can include four-form fluxes $G_4$ as part of 
the background, which preserve the Calabi-Yau metric up to warping \cite{Becker:1996gj}. Such fluxes 
give a potential to the complex structure moduli at tree level, which can be 
expressed in the resulting three-dimensional $\mathcal{N}=2$ theory in terms of the 
Gukov-Vafa-Witten (GVW) superpotential~\cite{Gukov:1999ya}
\begin{equation}\label{GVWsup}
  W_{\rm GVW} = \int_X G_4\wedge \Omega \,.
\end{equation}
The minima of the induced scalar potential are solutions of the F-term 
equations $D_IW=0$, $I=1,...,h^{1,3}(X)$. They are supersymmetric Minkowski 
vacua if furthermore $W_{\rm GVW}=0$. This implies that the complex structure 
must be such that $G_4 \in H^{2,2}(X)$.  It is commonly believed that a typical 
$G_4$ flux fixes all of the complex structure moduli. The argument for this is 
simple: there are as many constraints as there are complex structure moduli. 
The implicit assumption which enters this argument is that each of the F-term 
equations is linearly independent, which is expected to hold for a `generic' 
choice of $G_4$. 

As a consequence of flux quantization \cite{Witten:1996md}, which says that 
$G_4+ \frac{c_2(X)}{2} \in H^4(X,\mathbb{Z}) $, sensible choices of $G_4$ form a 
lattice, which begs the questions what precisely is meant by a `generic' choice 
of flux in this context. Complicating matters even more, there is the 
consistency condition commonly refereed to as $M2$-tadpole 
cancellation~\cite{Becker:1996gj}, which bounds the length squared of possible 
flux choices from above. Although it is always possible to find lattice vectors 
such that all F-term equations become linearly independent, this might require 
to pick lattice sites which are far away from the origin and hence too long to 
satisfy the bound imposed by the $M2$-tadpole.\footnote{There is also the possibility that the F-term equations have no solutions, as explained in \cite{Braun:2008pz} for M-theory compactifications on $K3\times K3$.}
The relevant question is hence: 
`is there a choice of flux such that all F-term equations are independent and 
the bound imposed by $M2$-tadpole cancellation is satisfied ?'.

This is a difficult question to study in general, and it may well be that 
the tadpole constraint has a strong selective power. This observation becomes 
particularly interesting when the fourfold $X$ is elliptically fibered and used 
as an F-theory background. In such compactifications, the four-dimensional gauge sector is 
engineered by appropriate singularities of $X$, and (part of) the complex 
structure moduli space of $X$ corresponds e.g. to adjoint Higgs fields. Complex 
structure moduli that do not receive a potential from \eqref{GVWsup} hence give 
rise to flat directions in the gauge sector, and the inability to stabilize all 
complex structure moduli corresponds to such flat directions inevitably being 
present\footnote{Some of the open string moduli may 
sit in matter multiplets that must be massless at the classical level to be 
consistent with some phenomenological requirements.}. On the other 
hand, loci of enhanced gauge symmetry are typically at very high codimension in 
the moduli space \cite{Braun:2014xka,Braun:2014lwp,Watari:2015ysa} and it would 
be a fascinating scenario if the consistency conditions only allowed fluxes 
that would select such loci for us \cite{Braun:2014ola,Braun:2014xka}.

The difficulty in working through explicit examples to shed light on the issues 
sketched above is mainly a technical one. Among the (known) Calabi-Yau 
fourfolds, a typical number of complex structure moduli is of the order of 
1000s. Evaluating \eqref{GVWsup} then requires to solve Picard-Fuchs equations 
of a ridiculously high degree. Furthermore, it is in general highly non-trivial 
to identify which elements of $H^4(X)$ are integral, so that they can be used 
to define an appropriately quantized flux. One method to find such a 
basis is given by mirror symmetry \cite{Bizet:2014uua}.

The main motivation of the present work is to further explore an alternative 
approach. The crucial idea underlying this approach is as follows: 
at supersymmetric Minkowski vacua, the properly quantized flux must be an 
element of $H^{2,2}(X) \cap H^4(X,\mathbb{Z})$ up to a 
shift $\tfrac12 c_2(X)$. The group $H^{2,2}(X) \cap H^4(X,\mathbb{Z}) \equiv 
H_{Hodge} (X)$ of Hodge cycles is not constant throughout complex structure moduli space, but 
may be enhanced at specific loci, called Hodge 
loci. This is analogous to the enhancement of the Picard lattice of K3 surfaces 
at Noether-Lefschetz loci. If we identify such a locus and switch on a flux 
which is proportional to one of the Hodge cycles appearing there, the model 
cannot be deformed away from this locus, as the flux is only of type $(2,2)$ on 
the Hodge locus, so that the associated F-term equations are necessarily 
violated away from it. Instead of picking a flux in $H^4(X)$ and asking where it 
drives the model, the strategy we want to use is to \emph{identify loci in the moduli 
space where supersymmetric fluxes are possible, and then ask if we can find a 
flux that traps it there}. See \cite{Aspinwall:2005ad} for 
a beautiful exposition of this idea. 

For K3 surfaces, the Torelli theorem implies that demanding for a single 
lattice vector in $H^2(K3,\mathbb{Z})$ to be in $H^{1,1}(K3)$ fixes one complex 
structure modulus. This is not true for fourfolds, where the number of complex 
structure moduli we need to tune for a single element $\eta$ in 
$H^4(X,\mathbb{Z})$ to be in $H^{2,2}(X)$ depends on both $X$ and $\eta$. Fixing 
all complex structure moduli then corresponds to finding a so-called `general' 
Hodge cycles for which the associated Hodge locus is just a point in the 
complex structure moduli space of $X$. If such a cycle furthermore satisfies 
the $M2$-tadpole constraint (after adding the piece $\frac{c_2(X)}{2}$), there is a 
$G_4$-flux that stabilizes all complex structure moduli. 

In order to identify Hodge cycles and their Hodge loci, we will make use of 
algebraic cycles of complex dimension two. These are Poincar\'e dual to forms of 
Hodge type $(2,2)$ and it is not hard to find instances which only appear at 
special loci in the moduli space. Such an approach was followed in 
\cite{Braun:2011zm}, and we will extend this work in several aspects. In 
\cite{Braun:2011zm}, the number of stabilized moduli was simply counted by 
working out how many polynomial deformations are frozen by the existence of a 
given algebraic cycle. As this tacitly assumes the validity of a version of the 
Hodge conjecture, such a method is insufficient for a reliably counting. This 
point which was adressed in \cite{Braun:2011zm} by using the relationship of 
complex structure moduli of F-Theory compactifications to open string moduli in 
IIB orientifolds, a way of reasoning that is not available for general M-Theory 
backgrounds on Calabi-Yau fourfolds. Furthermore, one may consider fluxes which 
are Poincar\'e dual to some linear combination of algebraic cycles. In this 
instance, studying polynomial deformations is simply not powerful enough to 
detect all flat directions. 

Working with the sextic fourfold $X_6$ at the Fermat point as a simple example, 
we show how to address both of these issues by directly evaluating the rank of 
the matrix
\begin{equation}
G_{IJ} \equiv \int_X G_4 \wedge D_I D_J \Omega \:,
\end{equation}
which counts the number of fixed complex structure moduli. The crucial 
ingredient needed to evaluate these integrals are the periods of variations of 
$\Omega$ over algebraic cycles, which have been computed for the sextic fourfold 
at the Fermat point in \cite{movasati_loyola_17,2018arXiv181203964V}. For the 
simplest class of algebraic cycle we show how to recover the periods (up to 
overall normalization) by exploiting the automorphism group of $X_6$, and 
construct fluxes that stabilize all complex structure moduli. These fluxes, 
however, significantly overshoot the tadpole constraint originating from the 
cancellation of $M2$-brane charge. Although a computation that confirms this in 
some form of generality is computationally too demanding to be within the scope 
of the present work, we take this as evidence for the tension between the $M2$ 
tadpole cancellation constraint and the desire to stabilize all complex structure moduli. 

As a further application we consider the interplay between fluxes and 
symmetries. In \cite{Giryavets:2003vd,Denef:2004dm} it was suggested to use fluxes respecting some symmetries of the complex structure moduli space, in order to stabilize all moduli. The trick is that one needs to solve only the F-term equations of the invariant moduli, as  the (many) F-terms of non-invariant complex structure deformations 
automatically vanish at a symmetric point. 
However, the argument does not take into account possible flat directions. In fact, 
we show that such flat directions are typically present in such setups. 
We give an example for the Fermat sextic fourfold.  
This  shows that caution has to be taken in using the trick of turning on symmetric fluxes to claim full complex structure moduli stabilization.

After reviewing some aspects of flux compactification in M-Theory on Calabi-Yau 
fourfolds in Section \ref{sect:background}, we discuss algebraic cycles at the 
Fermat point of the sextic fourfold in Section 
\ref{Sec:FermatSextAlgFluxes}. In Section \ref{Sect:residues}, we
describe the middle cohomology of $X$, the span of algebraic cycles, and 
variations of the holomorphic top-form $\Omega$ using residues of holomorphic 
forms with poles on $\P^5$. Some technical background on residues and rational 
forms are contained in an appendix. After introducing expressions for periods of 
residue forms on algebraic cycles, we apply these to several examples in Section 
\ref{sect:fluxstab}, and give some estimates that quantify the tension between 
complete moduli stabilization and tadpole cancellation. Moduli stabilization in 
the presence of fluxes respecting a symmetry is disussed in Section~\ref{sect:symmetry_actions}. We close with a discussion of open issues and 
future directions.

\section{Fluxes and Moduli Stabilization}\label{sect:background}

In this paper we consider M-theory compactified on a CY fourfold $X$. The 
resulting low energy theory is a three dimensional (3d) $\mathcal{N}=2$ 
supergravity, i.e. a theory with four supercharges. The metric deformations 
preserving the Calabi-Yau condition are called metric moduli and become massless 
scalars in the 3d theory. For CY fourfolds $X$, the metric moduli are encoded in 
the $h^{1,1}(X)$ periods of the K\"ahler form $J$ and the $h^{1,3}(X)$ 
independent deformations of the holomorphic $(4,0)$-form $\Omega$. 
These moduli are called \emph{K\"ahler moduli} and \emph{complex structure 
moduli}, respectively.  There are also $h^{1,1}(X)$ axionic moduli that come 
from the dimensional reduction of the eleven-dimensional (11d) sugra seven-form $C_7$ (the dual 
of $C_3$), which complexify the \emph{K\"ahler moduli}.

The dynamics of the moduli is determined by the K\"ahler potential
\be\label{Kahlerpottot}
 K = K_{c.s.} + K_{K}\:,
\ee
with 
\be 
 K_{c.s.} = -\ln\left( \int_X \Omega\wedge\bar{\Omega}  \right) \qquad\mbox{and}\qquad 
 K_K = -3 \ln \left( \frac{1}{4!} \int_X J\wedge J\wedge J\wedge J \right) \:.
\ee

One can switch on a non-zero vev for the four-form flux $G_4=dC_3$, that is quantized according to 
\be
 G_4 + \frac{c_2(X)}{2} \in H^4(X,\mathbb{Z}) \:.
\ee
where  $c_2(X)$ is the second Chern class of the tangent bundle of $X$.

A non zero flux along internal directions generates a potential for the metric 
moduli after compactification \cite{Haack:2001jz}. This can be understood 
from the 11d $C_3$ kinetic term $\int G_4\wedge\ast G_4$, which depends on the 
metric through the Hodge star operator $\ast$). The minima of the supergravity 
scalar potential are given by the solutions of the following equations
\begin{equation}\label{MinEqns}
\left\{\begin{array}{lccl}
 D_I W = 0 &&& I=1,...,h^{3,1} \\   \\
 \partial_k \tilde{W} = 0 &&& k=1,...,h^{1,1} 
\end{array}\right.  
\end{equation}
where 
\be 
W = \int_X G_4\wedge \Omega  \qquad\mbox{and}\qquad 
\tilde{W}= \int_X G_4\wedge J\wedge J \,.
\ee
Here $W$ is the GVW superpotential \cite{Gukov:1999ya} and 
$D_I = \partial_I+\partial_IK$, with $K$ the K\"ahler potential 
\eqref{Kahlerpottot}. The index $I$ runs over the complex structure moduli, and 
the index $k$ runs over the K\"ahler moduli.

These minima are at zero cosmological constant (i.e. they are Minkowski 
vacua). They are furthermore  supersymmetric if the vev of $W$ vanishes, 
i.e. $W|_{\rm min}=0$.  This condition together with \eqref{MinEqns} can be 
rephrased by saying that the four-form flux must lie in $H^{2,2}_{\rm prim}(X)$, 
i.e.\emph{ $G_4$ must be a primitive four-form of Hodge type $(2,2)$}. We now 
explain this. The same can be done in the dual type IIB compactification on CY 
orientifolds \cite{Giddings:2001yu}, see also 
\cite{Denef:2008wq} for an overview over the classic literature on the subject.

We first explain why $G_4$ is of Hodge type $(2,2)$: 
\begin{itemize}
\item The condition $W=0$ means $$\int_X G_4\wedge \Omega = 0;$$ this implies that the $(0,4)$ component of $G_4$ vanishes. Since $G_4$ is real, also its $(4,0)$ component is zero.
\item The condition $D_IW=0$ means $$\int_X G_4\wedge D_I\Omega = 0 
\qquad\forall I;$$ since the forms $D_I\Omega$ give a basis of $H^{3,1}(X)$ 
\cite{Strominger:1990pd,Greene:1993vm}, the $(1,3)$ and $(3,1)$ components of 
$G_4$ vanish. \end{itemize}
We then see that only the $(2,2)$ part of $G_4$ survives. 

As regarding the primitivity condition, expand first the K\"ahler form $J$ in a 
basis of harmonic $(1,1)$-forms $\omega_k$: $J=t^k\omega_k$. $t^k$ are the 
$h^{1,1}(X)$ K\"ahler moduli. The condition $\partial_k\tilde{W}=0$ means
$$
\int_X G_4\wedge J \wedge \omega_k = 0 \qquad\forall k
$$
that implies $G_4\wedge J=0$, i.e. $G_4$ is a primitive form.

When the flux, as required, belongs to $H^{2,2}_{\rm prim}(X)$, then it is 
also \emph{self-dual}, i.e. $\ast G_4 = G_4$. This, in particular, implies that 
the contribution of $G_4$ to the M2-charge, i.e.
\be 
Q_{M2}^{\rm flux} = \frac12 \int_X G_4\wedge G_4 \:,
\ee
is positive definite. In order to be possible to satisfy the M2-tadpole 
cancellation condition, 
\begin{equation}
Q_{M2}^{\rm flux} + N_{M2} =  \frac{\chi(X)}{24}\,
\end{equation}
without introducing anti-branes, $Q_{M2}^{\rm flux}$ must be smaller than the 
contribution coming from the geometry, i.e. $Q_{M2}^{\rm flux}\leq 
\frac{\chi(X)}{24}$. 

Let us now concentrate on the complex structure moduli. We choose a point in 
the complex structure moduli space that satisfies $D_IW=0$ and $W=0$. We take 
coordinates $s_I$ such that this point is at $ \mathbf{s}= (s_0,s_1,s_2, ...) = 
0 $. The holomorphic $(4,0)$-form at a generic point is $\Omega(s)$ and 
$W(s)=\int_X G_4 \wedge  \Omega(s)$. We then have 
\be \label{DIWatMin}
\left. D_I W(s)  \right|_{s=0} = 0 \:.
\ee
A flat direction of the potential is a curve $s(t)$ in the moduli space passing 
through $s=0$ at $t=0$ that satisfies the minimum condition for all $t$ in a 
neighborhood of $t=0$, i.e.  
\be \label{FlatDirecGenLarge}
 D_I W(s(t))  = 0 \qquad\qquad \forall t  \:. 
\ee
Expanding around $t=0$ and keeping the leading term at small $t$,  one finds the infinitesimal expression for \eqref{FlatDirecGenLarge}, i.e.
\be \label{FlatDirecGen}
\dot{s}_J(0) \partial_J D_I W(0)  = 0 \:.
\ee
Notice that $\partial_J D_I W(0) =D_J D_I W(0) $, since the two expressions 
differ by $(\partial_JK ) D_I W(0)$ which vanishes because of 
\eqref{DIWatMin}.\footnote{ In general $D_IW$ is not holomorphic, as there is 
the term $(\partial_I K) W$ with $\partial_I K$ non-holomorphic in the complex 
structure moduli chiral superfields. Hence one may expect in 
\eqref{FlatDirecGen} a term involving $\bar{\partial}_J D_I W$. However the 
extra term is $(\partial_I \partial_{\bar{J}} K) W$, which vanishes at 
$s=0$ because of the $W=0$ condition.
}
The vectors $\dot{s}_J(0)$ solving \eqref{FlatDirecGen} give the flat directions.

We hence conclude that in order to have no flat directions at $s=0$ we need that the matrix
\begin{equation}\label{eq:matrixfluxrank}
G_{IJ} :=    \left. D_J D_I W(s)\right|_{s=0}    
\end{equation}
has maximal rank.
More generally, the rank of the matrix \eqref{eq:matrixfluxrank} counts how many complex structure moduli are stabilized by $G_4$. 
This is called the codimension of the Hodge locus of $G_4$ in math literature, see \cite{voisin2010hodge,movasati_book} for a review.

The Poincar\'e dual of an algebraic four-cycle is a four-form of 
type $(2,2)$. When the fourfold is at a specific point in the complex structure 
moduli space, one may be able to construct explicit algebraic cycles, as we will 
do for the sextic fourfold. One can then use them  to define a choice of 
properly quantized flux that is a primitive $(2,2)$-form at that specific point. 
The question we want to address here, is how many moduli are stabilized once such 
a flux is introduced: any deformation that breaks originates in a $G_4$ flux 
not purely of type $(2,2)$ is lifted by the flux potential.

Let us come back to the GVW superpotential that generates the minima condition 
for the complex structure moduli. The part of the flux $G_4$ that contributes to 
the superpotential, the F-term conditions and the stability condition is the one 
that has non-zero intersection with $\Omega(s)$ and its derivatives. 
Here by `intersection' we mean the product given by the inner form $a_1\cdot 
a_2\equiv \int_X a_1\wedge a_2$. The holomorphic four-form and its derivatives do 
not span the full middle cohomology $H^4(X)$, but only the primary horizontal 
subspace \cite{Strominger:1990pd,Greene:1993vm}. In contrast, forms of 
Hodge type $(2,2)$ defined by intersections of divisors lie in the primary 
vertical subspace, which is perpendicular to the primary horizontal 
subspace\footnote{In general, there can be directions in $H^{2,2}(X)$ which do 
not lie in either subspace \cite{Braun:2014xka}.}. To study stabilization of 
complex structure moduli we hence need to consider $G_4$ fluxes that lie in the 
horizontal subspace of $H^4(X)$ (apart from the piece $\tfrac12 c_2(X)$ that is 
forced on us by quantization). The algebraic cycles we consider here are 
exactly of this type \cite{Braun:2011zm}.

\section{Fermat Sextic Fourfold and Algebraic Cycles} 
\label{Sec:FermatSextAlgFluxes}

The manifold of interest to us in this paper is the sextic fourfold. 
A sextic fourfold at a generic point in its moduli space is defined by the 
vanishing of a homogeneous polynomial of degree $6$ in $\P^5$:
\begin{equation}
\mathcal{X}_6: \hspace{.2cm} x_0^6 + x_1^6 + x_2^6 + x_3^6 + x_4^6 + x_5^6 + \sum_\mathbf{a} c_\mathbf{a} \prod x_i^{a_i} = 0 \,  
\end{equation}
where $\mathbf{a} = (a_1,\cdots ,a_5)$ are integers 
such that $\sum a_i = 6$ and the $c_\mathbf{a}$ are complex coefficients that 
can be thought of as deformations of the complex structure. The topological 
numbers of $\mathcal{X}_6$ are 
\begin{equation}
h^{1,1}(\mathcal{X}_6) = 1 \hspace{1cm} h^{2,1}(\mathcal{X}_6) = 0 \hspace{1cm} h^{3,1}(\mathcal{X}_6) = 426 \hspace{1cm} h^{2,2}(\mathcal{X}_6) = 1752\, . 
\end{equation}
It follows that $\chi(\mathcal{X}_6) = 2610$ and $b^4_+(\mathcal{X}_6) = 1754$, 
$b^4_-(\mathcal{X}_6) = 852$. 

The single class in $h^{1,1}(\mathcal{X}_6)$ is generated by the restriction of 
the hyperplane class $H$ of $\P^5$, and any K\"ahler form on $X_6$ is 
necessarily proportional to $H$. There is a unique generator of the primary 
vertical subspace $H^{2,2}_V(\mathcal{X}_6)$ which is given by $H \cdot H \equiv 
H^2$ and which is always proportional to the square of the K\"ahler form. 

The orthogonal directions to $H^2$ in $H^4(\mathcal{X}_6)$ are hence all 
primitive, i.e. $h^{2,2}_{\mbox{\tiny prim}}(\mathcal{X}_6) = 1751$, and can 
be shown\footnote{We will see this explicitly in Section 
\ref{sect:csmoduliresidues}. This can also be shown by computing that the 
dimension of the primary vertical subspace of the mirror, 
$h^{2,2}_V(X^\vee)=1751$ and using that primary vertical and horizontal 
subspaces are swapped by the mirror map. } to all belong to the primary 
horizontal subspace $H^{4}_H(\mathcal{X}_6)$, which has dimension $1751$. The 
second Chern character of $\mathcal{X}_6$ is
\begin{equation}
 c_2(\mathcal{X}_6) = 15 H^2 \, . 
\end{equation}
The term $\tfrac12 c_2(\mathcal{X}_6)$ is hence not integral, so that 
flux quantization forces us to include a half-integral flux proportional to 
$H^2$.

For a typical choice of the $c_\mathbf{a}$, the only algebraic cycles contained 
in $\mathcal{X}_6$ are complete intersections of $\mathcal{X}_6$ with multiples 
of the hyperplane divisor in $\P^5$. On $\mathcal{X}_6$ the classes of these are
proportional to $H^2$, so that the rank of $H^{2,2}(\mathcal{X}_6) \cap 
H^4(\mathcal{X}_6)_{prim}$ is zero. As $H^2$ is never primitive, there are 
furthermore no supersymmetric fluxes along this direction. If we tune the 
$c_\mathbf{a}$ to special values, the situation changes and the rank of 
$H^{2,2}(\mathcal{X}_6) \cap H^4(\mathcal{X}_6,\mathbb{Z})_{prim}$ becomes 
non-zero. 

Let us hence make a specific choice and set all $c_\mathbf{a} = 0$, 
which puts us on the Fermat point\footnote{This is also called the Gepner 
point (if one thinks in terms of the worldsheet CFT of strings propagating on 
$\mathcal{X}_6$) or the Brieskorn-Pham point (if one things in terms of 
singularity theory). } of the moduli space of the 
sextic. We will denote the sextic fourfold at the Fermat point by:
\begin{equation}\label{eq:sextic4fold}
X_6: \hspace{.2cm} x_0^6 + x_1^6 + x_2^6 + x_3^6 + x_4^6 + x_5^6 = 0 \, .
\end{equation}
As the above equation describes a smooth submanifold of $\P^5$, the topological 
numbers of the Fermat sextic are the same as those of $\mathcal{X}_6$. Only
the group $H^{2,2}(X_6) \cap H^4(X_6)_{prim}$ is different from the case of a 
generic sextic fourfold: it has the maximal possible rank of $1751$.

\subsection{Algebraic cycles at the Fermat point}

It is not hard to find the simplest type of algebraic cycle sitting inside
$X_6$. Take e.g. $x_0 = \alpha x_1$, $x_2 = \beta x_3$ and $x_4 = \gamma x_5$
for $\alpha^6 = \beta^6 = \gamma^6 = -1$. In this case, these three equations imply 
\eqref{eq:sextic4fold}, so that they define a subvariety of complex codimension $3$ inside $\P^5$, which is complex codimension $2$ in $X_6$. 

Using the large group of automorphisms of $X_6$, we can immediately write 
down the general form of such cycles as
\begin{equation}\label{eq:cjkmsexticv2}
\begin{aligned}
C_{\mathbf{\sigma}}^{\,\boldsymbol{\ell}}:\,\, x_{\sigma(0)} =  e^{i \pi/6} 
e^{i \pi \ell_0/3} x_{\sigma(1)} &\hspace{.3cm}& x_{\sigma(2)} =  e^{i \pi/6} 
e^{i \pi \ell_1/3} x_{\sigma(3)}  &\hspace{.3cm}& x_{\sigma(4)} =  e^{i \pi/6} 
e^{i \pi \ell_2/3} x_{\sigma(5)}   \\
\end{aligned}
\end{equation}
Here the $\ell_i \in \{0,1,2,3,4,5\}$ specify which sixth root of unity we are 
using and $\sigma$ is a permutation of $\{0,1,2,3,4,5\}$ which specifies which 
coordinates 
are paired to form $C_{\mathbf{\sigma}}^{\,\boldsymbol{\ell}}$. 

The existence of such algebraic cycles can also be inferred by writing the 
defining equation of $X_6$ \eqref{eq:sextic4fold} in the following 
`factorized' form
\be
\begin{aligned}
\prod_{\ell_0=0}^5\left( x_{\sigma(0)} - e^{i\pi/6}e^{i\pi \ell_0/3} x_{\sigma(1)} \right) + 
\prod_{\ell_1=0}^5\left( x_{\sigma(2)} - e^{i\pi/6}e^{i\pi \ell_1/3} x_{\sigma(3)} \right)   \\ + 
\prod_{\ell_2=0}^5\left( x_{\sigma(4)} - e^{i\pi/6}e^{i\pi \ell_2/3} x_{\sigma(5)} \right)=0\: . 
\end{aligned}
\ee
This gives a hint of how other instances of algebraic cycles can be found. 
Another factorization of the defining equation for $X_6$ is:
\begin{eqnarray}
0 &=& \left( x_0^3 + e^{i\pi k} x_1^3 + i  x_2^3 \right) \left( x_0^3 + e^{i\pi k} x_1^3 - i  x_2^3 \right) + 
\prod_{m=0}^2 \left( x_3^2 - 2^{1/3}e^{i\pi k} e^{\frac{2\pi i m}{3}} x_0 x_1 \right) \nonumber \\
 && +\,\,  \prod_{\ell=0}^6\left( x_{4} - e^{i\pi/4}e^{i\pi \ell/2} x_{5} \right)
\end{eqnarray}
up to permutation of the four coordinates and for $k=0,1$. One then realizes the existence of the algebraic cycles
\begin{equation}\label{eq:Ckjmell}
C_\sigma^{kjm\ell} :\,\,\,\,
\begin{aligned}
x_{\sigma(0)}^3 + e^{i\pi k} x_{\sigma(1)}^3 + i e^{i \pi j} x_{\sigma(2)}^3 = 0 \\
x_{\sigma(3)}^2 - 2^{1/3} e^{i\pi k} e^{ \frac{2\pi i m}{3} }x_{\sigma(0)} x_{\sigma(1)} =0 \\
x_{\sigma(4)} -  e^{i \pi/6} e^{i \pi \ell/3} x_{\sigma(5)} =0
\end{aligned}
\end{equation}
for $k,j \in \Z/2\Z$, $m \in \Z/3\Z$ and $\ell\in \Z/6\Z$. 
Note that $x_{\sigma(4)}$ and $x_{\sigma(5)}$ are paired in a similar way as 
before, whereas a more complicated factorization is used for the remaining four 
coordinates. These cycles are a lift of the cycles that were used to construct 
the N\'eron-Severi group of Fermat sextic surfaces in \cite{Aoki1983}, where 
famously using only lines is no longer sufficient 
\cite{shioda1982picard}. 

An example of a completely non-linear factorization of \eqref{eq:sextic4fold} 
is given by
\begin{eqnarray}
0&=& \prod_{s=0}^2 \left( x_0^2 + e^{\frac{2\pi i}{3}(k_1+s)}x_1^2 + e^{\frac{2\pi i}{3}(k_2+2s)}x_2^2  \right) 
+ \prod_{s=0}^2 \left( x_3^2 + e^{\frac{2\pi i}{3}(k_4+s)}x_4^2 + e^{\frac{2\pi i}{3}(k_5+2s)}x_5^2  \right) \nonumber \\
	&& + \,\,3\prod_{n=0}^1 \left(  i \,e^{i\pi n}e^{\frac{i\pi}{3}(k_1+k_2)}x_0x_1x_2 + \,e^{\frac{i\pi}{3}(k_4+k_5)}x_3x_4x_5  \right)
\end{eqnarray}
for some $k_1,k_2,k_4,k_5\in \mathbb{Z}/3\mathbb{Z}$.
We then find that the Fermat sextic fourfold contains the algebraic cycles
\begin{equation}\label{eq:Ck1k2k4k5n}
C_\sigma^{k_1 k_2 k_4 k_5 n}: \,\,
\begin{aligned}
x_{\sigma(0)}^2 + e^{\frac{2\pi i}{3}k_1}x_{\sigma(1)}^2 + e^{\frac{2\pi i}{3}k_2}x_{\sigma(2)}^2 =0 \\
x_{\sigma(3)}^2 + e^{\frac{2\pi i}{3}k_4}x_{\sigma(4)}^2 + e^{\frac{2\pi i}{3}k_5}x_{\sigma(5)}^2 =0 \\
 i \,e^{i\pi n}e^{\frac{i\pi}{3}(k_1+k_2)}x_{\sigma(0)}x_{\sigma(1)}x_{\sigma(2)} + \,e^{\frac{i\pi}{3}(k_4+k_5)}x_{\sigma(3)}x_{\sigma(4)}x_{\sigma(5)} = 0 
\end{aligned}\,\,\, ,
\end{equation}
with $k_i \in \Z/3 \Z$ and $n \in \Z/2 \Z$. 

There are further algebraic cycles of the form $f_0=f_1=f_2=0$ contained in 
$X_6$ which can be seen by constructing other factorizations of the form
\begin{equation}\label{eq:factorization_for_cialgcycle}
X_6: f_0 P_0 + f_1 P_1 + f_2 P_2 = 0 \,, 
\end{equation}
see \cite{aoki1987_alg_cycles} and \cite{movasati_book} for a more systematic 
treatment. Note also that all of the examples of algebraic cycles we have given 
are complete intersections inside the ambient $\P^5$, which points to another 
direction of generalization: algebraic cycles which are not complete 
intersections. Over $\mathbb{Q}$, it is known, however, that all of 
$H^{2,2}_{prim}(X_6)$ is generated by the above algebraic cycles 
\cite{shioda_hodge_conj_fermat_79,Aoki1983,aoki1987_alg_cycles}.

\subsection{Some properties of algebraic cycles}

Having introduced some algebraic cycles on the Fermat sextic, let us study some
of their properties. We will limit our discussion mostly to the `linear' 
algebraic cycles $C_{\mathbf{\sigma}}^{\,\boldsymbol{\ell}}$. 

As each of the $C_{\mathbf{\sigma}}^{\,\boldsymbol{\ell}}$ is given by three 
linear 
equations inside $\P^5$, each such cycle has the topology of $\P^2$. To 
compute intersection numbers, we can use the following trick. Consider a 
complete intersection of $X_6$ with $x_0 - \alpha x_1=0$ for $\alpha^6=-1$ and 
$x_2 - \beta x_3=0$ for $\alpha^6=\beta^6-1$. The resulting cycle on $X_6$ is in 
the class $H^2$ restricted to $X_6$. Using \eqref{eq:sextic4fold}, however, we 
see that this cycle is reducible into a sum of six of the 
$C_{\mathbf{\sigma}}^{\,\boldsymbol{\ell}}$. We may hence write
\begin{equation}\label{eq:sumrulelincycles}
H^2 = \sum_{\ell_0=0}^5 C_{\mathbf{\sigma}}^{\,\boldsymbol{\ell}} 
\end{equation}
for any choice of $\sigma$ and every $\ell_1$ and $\ell_2$. As $H^2 \cdot H^2 = 
6$ on $X_6$ and $H^2 \cdot C_{\mathbf{\sigma}}^{\,\boldsymbol{\ell}}$ is the 
same for every $C_{\mathbf{\sigma}}^{\,\boldsymbol{\ell}} $ by symmetry, it 
follows that
\begin{equation}
 H^2 \cdot C_{\mathbf{\sigma}}^{\,\boldsymbol{\ell}}  = 1\, . 
\end{equation}
Using the observation that 
$C_{\mathbf{\sigma}}^{\,\boldsymbol{\ell}} \cdot 
C_{\mathbf{\sigma}}^{\,\boldsymbol{\ell}'}=0$ if 
$\boldsymbol{\ell}$ and $\boldsymbol{\ell}'$ differ in all three components 
(together with a similar rule when intersection algebraic cycles employing 
different permutations $\sigma$), the above can be iterated to find that the 
intersection numbers follow the 
pattern
\begin{equation}
\begin{array}{c|c}\label{eq:int_numbers_lin_cycles}
 \mbox{dim} (C_{\mathbf{\sigma}}^{\,\boldsymbol{\ell}} \cap 
C_{\mathbf{\sigma}'}^{\,\boldsymbol{\ell'}}) &  
C_{\mathbf{\sigma}}^{\,\boldsymbol{\ell}} \cdot 
C_{\mathbf{\sigma}'}^{\,\boldsymbol{\ell'}} \\
 \hline 
2 & 21 \\
1 & -4 \\
0 & 1 \\
\emptyset & 0
\end{array}\, .
\end{equation}
i.e. the dimension of the intersection of two algebraic cycles in $\P^5$ 
determines the intersection number between the associated homology 
classes\footnote{If two algebraic cycles do not intersect 
transversely, there is always a pair of homologous (typically non-holomorphic) 
cycles that do intersect transversely.}. For any pair of permutations $\sigma$ 
and $\sigma'$, the intersection numbers can also be expressed in terms of 
relations on the $\ell_i$ and $\ell_j'$.

Although one can work out the details using the same approach, such a simple 
pattern is not obeyed by the other algebraic cycles introduced in the last 
section. Self-intersections of any algebraic cycle $C_{f_0,f_1,f_2}$ of 
complete 
intersection type given by $f_0=f_1=f_2=0$ can however be worked out using 
adjunction, and the result is \cite{movasati_book}
\begin{equation}
C_{f_0,f_1,f_2} \cdot C_{f_0,f_1,f_2}  = d_0 d_1 d_2 (36 - 6 (d_0 + d_1 + d_2) + 
d_0 d_1 + d_0 d_2 + d_1 d_2 )
\end{equation}
where $d_i$ are the degrees of the polynomials $f_i$.

\subsection{Algebraic cycles and their Hodge loci}

Let us now try to see how many moduli we expect to be stabilized by demanding 
that any of the cycles $C_{\mathbf{\sigma}}^{\,\boldsymbol{\ell}}$ remains of 
type 
$(2,2)$. \footnote{\label{footnote3} Straightforwardly using 
$C_{\mathbf{\sigma}}^{\,\boldsymbol{\ell}}$ as a flux is at odds with 
primitivity and 
flux quantization. To achieve a primitive flux, we would need to choose $G_4 = 
C_{\mathbf{\sigma}}^{\,\boldsymbol{\ell}} - \tfrac16 H^2$. This is at odds with 
flux 
quantization, which requires $G_4$ to be integral up to a piece $\frac12 H^2$. 
One would hence need to consider $G_4 = 3 
C_{\mathbf{\sigma}}^{\,\boldsymbol{\ell}} - 
\tfrac12 H^2$. Any piece proportional to $H^2$ does however not influence 
complex structure deformations, and the number of stabilized moduli is the same 
for $C_\sigma^\ell$ and $3C_\sigma^\ell$, so that we prefer to simply ask about 
the `Hodge Locus' of $C_{\mathbf{\sigma}}^{\,\boldsymbol{\ell}}$ here.} We can 
work out 
the number of polynomial deformations which are obstructed by demanding that 
$C_{\mathbf{\sigma}}^{\,\boldsymbol{\ell}}$ is an algebraic cycle as follows. 
First one 
observes that it does not matter which 
$C_{\mathbf{\sigma}}^{\,\boldsymbol{\ell}}$ we are 
talking about as they are all equivalent modulo automorphisms of the Fermat 
sextic. 
For $\alpha^6 = -1$, let us hence consider the cycle 
\be 
C\,: \qquad x_0- \alpha x_1 =0\,, \qquad x_2- \alpha x_3 =0\,, \qquad x_4- 
\alpha x_5=0 \:.
\ee
We can introduce a new set of coordinates:
\be 
 (y_0,y_1,y_2,y_3,y_4,y_5)=(x_0-\alpha x_1,x_0+\alpha x_1,x_2-\alpha 
x_3,x_2+\alpha x_3,x_4-\alpha x_5,x_4+\alpha x_5)\:, 
\ee
in terms of which the Fermat sextic equation \eqref{eq:sextic4fold} becomes
\be
 y_0 y_1 (3y_0^4+10y_0^2y_1^2+3y_1^4) + y_2 y_3(3y_2^4+10y_2^2y_3^2+3y_3^4) + 
y_4 y_5(3y_4^4+10y_4^2y_5^2+3y_5^4) = 0
\ee
Polynomial deformations are counted by counting monomials of degree 6 modulo the 
Jacobi ideal.\footnote{This way of counting deformations is explained in some 
more detail in Section \ref{Sect:residues}. It gives the same result as counting 
monomials modulo automorphism of $\mathbb{P}^5$, but is more convenient here.}  
There are $\binom{11}{6}=462$ possible monomials of degree 6 in 6 variables. The 
Jacobi ideal is generated by 
\be
(3y_{2\kappa+1}^5+30y_{2\kappa+1}^3y_{2\kappa}^2+15y_{2\kappa+1}y_{2\kappa}^4,  
3y_{2\kappa}^5+30y_{2\kappa}^3y_{2\kappa+1}^2+15y_{2\kappa}y_{2\kappa+1}^4)  
\,\,\, \mbox{with}\,\,\,\kappa=0,1,2\:.
\ee
We can use the Jacobi ideal to eliminate all monomials proportional to $y_i^5$, 
and there are $36$ such monomials. Hence the number of complex structure moduli 
is $426$, which equals $h^{3,1}(X)$ as expected. 

We now demand the cycle $C^{\boldsymbol{\ell}}_\sigma$ to persist as an 
algebraic cycle. This is the case only if the deformed fourfold is of the form 
\begin{equation}
 y_0 P_5(y_0,...,y_5) +   y_2Q_5(y_0,...,y_5) +   y_4R_5(y_0,...,y_5)=0
\end{equation}
where $P,Q,R$  are homogeneous polynomial of degree $5$. This means that we can 
use  only the monomials that have a factor of $y_0$, $y_2$ or $y_4$ to deform 
the Fermat sextic. 
The obstructed deformations are then monomials of degree 6 in the three 
coordinates $y_1,y_2,y_3$. This gives $\binom{8}{6}=28$ deformations. We have 
to 
subtract the $3\times 3=9$ monomials that are in the Jacobi ideal. We then 
obtain that $19$ moduli are fixed by demanding that any of the 
$C_\sigma^\ell$ persists as an algebraic cycle.

Again, there is a general version of this method that can be applied to any 
cycle $C_{f_0,f_1,f_2}$. The result only depends on the degrees of the 
polynomials $f_i$ and can be found in \cite{movasati_book}. 

The main issue with this approach which requires us to work harder is 
that we are interested in stabilizing all complex structure moduli, which 
forces 
us to consider linear combinations of algebraic cycles. Demanding that a single 
cycle $C_{\mathbf{\sigma}}^{\,\boldsymbol{\ell}}$ be algebraic only fixes some, 
but not 
all of the moduli. Similar results are obtained for other cycles 
$C_{f_0,f_1,f_2}$, so that we are led to consider linear combinations of (the 
Poincar\'e duals of) algebraic cycles. Merely counting polynomial deformations 
then becomes useless and we need a method to evaluate \eqref{eq:matrixfluxrank} 
in order to treat such situations.

A further issue that deserves some discussion concerns the Hodge conjecture. 
While the Hodge conjecture over $\mathbb{Q}$ has been proven for the Fermat 
sextic \cite{shioda_hodge_conj_fermat_79,aoki1987_alg_cycles} (see 
Section~\ref{Sect:residues} for more details), we do not know if it is true in 
general. This means for other points in the moduli space of the sextic or for 
other fourfolds, the number of polynomial deformations that are fixed by 
demanding a cycle stays algebraic may not equal the number of complex structure 
deformations that are fixed by demanding that the dual integral $(2,2)$ form 
stays of type $(2,2)$. Of course every algebraic cycle must be dual to an 
integral form of type $(2,2)$, but it is not clear that every integral form of 
type $(2,2)$ can be represented by a linear combination of algebraic cycles. In 
our context this implies that there might be extra flat directions that cannot 
be detected from polynomial deformations. Again, being able to evaluate 
\eqref{eq:matrixfluxrank} settles this issue.

\section{Residues of Rational Forms and Complex Structure 
Deformations}\label{Sect:residues}

In this section we will use the techniques of rational forms to explicitly 
describe the middle cohomology of $X$, some background on these techniques is 
given in Appendix~\ref{App:RationalForms}. This is then used to describe 
complex 
structure deformations and moduli stabilization for fluxes defined by (sums of) 
algebraic cycles. Throughout this section, $X$ is the Fermat sextic fourfold 
\eqref{eq:sextic4fold}. 

\subsection{Middle cohomology from residues of rational forms}

As reviewed in  Appendix~\ref{App:RationalForms}, primitive forms of Hodge 
type $(p,4-p)$ on the Fermat sextic are described as residues of rational forms 
\begin{equation}\label{eq:forms_in_PminusX}
		\varphi = \frac{P(x)}{Q(x)^{5-p}}\Omega_0\:,
\end{equation}
on $\P^5$. Here $Q=0$ is the hypersurface equation defining the Fermat sextic 
fourfold $X$, $\Omega_0$ is a fixed differential form on $\P^5$ that is 
completely antisymmetric in the homogeneous coordinates $x_i$, and $P$ is a 
homogeneous polynomial of degree $6(4-p)$. The residue map is linear, maps 
surjectively to the primitive forms in the middle cohomology of $X$, and 
becomes 
injective when restricting to polynomials $P$ which are not contained in the 
Jacobi ideal of $Q$.  

Let us apply these statements to reproduce the Hodge numbers of the sextic. To 
work out the dimensions of the rings we are going to consider, it is beneficial 
to know that there are
\begin{equation}
\#(\mbox{degree l in m variables})  = \binom{l+m-1}{l}
\end{equation}
terms in a homogeneous polynomial of degree $l$ in $m$ variables.

The existence of a unique $(4,0)$-form up to scaling follows from the fact that 
for $p=4$, $P$ is just a number. To find classes of $(3,1)$-forms, we hence 
need 
to consider the case 
$p=3$, i.e. homogeneous polynomials of degree $6$ modulo the Jacobi ideal of 
$Q$:
\begin{equation}\label{eq:h31primsextic}
H^{3,1}_{\mbox{\tiny prim}}(X) = \frac{\mathbb{C}[x_0,\cdots, 
x_{5}]_{6}}{\langle \partial_i Q \rangle}
= \frac{\C[x_0,\cdots,x_5]_{6}}{\langle x_0^5, \cdots, x_5^5 \rangle} \:.
\end{equation}
The ring of homogeneous polynomials of degree $6$ in $6$ variables has dimension 
$462$. At the Fermat point, the Jacobi ideal is generated by the polynomials 
$x_i^5$ for all $i$, so that $6 \cdot 6 =36$ generators of 
$\C[x_0,\cdots,x_5]_6$ are contained in the Jacobi ideal of $Q$. We hence 
recover the familiar number $h^{3,1}_{\mbox{\tiny prim}}(X) = h^{3,1}(X) = 426$. 
Finally, for $H^{2,2}_{\mbox{\tiny prim}}(X)$ we have $p=2$, so that we need to 
count polynomials of degree $12$ modulo the Jacobi ideal:
\begin{equation}
H^{2,2}_{\mbox{\tiny prim}}(X) = \frac{\C[x_0,\cdots,x_5]_{12}}{\langle x_0^5, 
\cdots, x_5^5 \rangle}\, . 
\end{equation}
We can work this out by noting that for each variable, the number of terms in 
$\C[x_0,\cdots,x_5]_{12}$ which are in the ideal $x_i^5$ is given by a 
homogeneous polynomial of degree $7$. Using this we need to take into account 
that for each pair of variables $x_i$ and $x_j$ there are terms $x_i^5 x_j^5 
P_2(x)$ for a polynomial $P_2(x)$ of degree $2$, which are in both the ideal 
generated by $x_i^5$ and $x_j^5$. Hence
\begin{equation}
\left|\frac{\C[x_0,\cdots,x_5]_{12}}{\langle x_0^5, \cdots, x_5^5 
\rangle}\right| =  \binom{17}{12} - 6 \cdot \binom{12}{7} + 15 \cdot\binom{7}{2} 
= 1751 = h^{2,2}_{\mbox{\tiny prim}}(X) \,.
\end{equation}

\subsection{Group actions and residues}\label{sect:group_on_residues}

The content of the last subsection can be rephrased by considering the natural 
group action of $G = \mu^{6}/\mu$ for $\mu = \Z/6\Z$ by coordinatewise 
multiplication:
\begin{equation}\label{eq:grouponfermat}
(x_0,\cdots,x_{d+1}) \rightarrow  (\zeta_0 x_0,\cdots,\zeta_{d+1} x_{d+1}) \
\end{equation}
for $\zeta_i^{n} = 1$ an $n$-th root of unity. The quotient arises because 
elements of $\mu^{6}$ for which all $\zeta_i$ are equal are inside the 
$\mathbb{C}^*$ acting on the homogeneous coordinates of $\P^{5}$.\footnote{Note 
that this group is larger than the group $\mu^4$ which is used in the 
Green-Plesser mirror construction: as we do not have a term proportional to 
$\prod_i x_i$ there is no need to impose $\prod_i \zeta_i= 1$.} 

Let us consider the character group $A$ of $G$, which is the group of 
representations by complex valued functions of $G$:
\begin{equation}
A = \{\a= (a_0,\cdots,a_{d+1})\,|\,\,  a_i \in \Z/6\Z \,\,\mbox{ and }\,\, 
\sum_i a_i =0 \,{\rm mod}\, 6 \} \:.
\end{equation}
The elements of $A$ are functionals on $G$ that associate to an element $g\in G$ 
the phase
\begin{equation}
\a(g) = \prod_i \zeta_i^{a_i} \, .
\end{equation}
Note that the condition $\sum_i a_i =0$ mod $6$ guarantees that the unit element 
of $G$ (i.e. $\zeta_0=...=\zeta_{d+1}$) is mapped to $1$, i.e. that this is 
indeed a group homomorphism.

The action of $G$ on $X$ induces an action on the middle cohomology 
$H^4(X,\C)$. One can use the character group $A$ to describe such an action. 
$G$ is an abelian group, so its elements can be diagonalized simultaneously. 
The elements $\a$ play the role of `eigenvalues'. We may define `eigencycles' 
relative to $\mathbf{a} \in A$ to be those classes $\eta$ for which 
\begin{equation}\label{eq:defeigencycle}
g^*\eta  = \a(g)\eta \hspace{1cm}  \forall g \in G \:. 
\end{equation}
For a given $\a$, we denote the span of the cycles which satisfy the above 
relation by $V(\a)$. 

The spaces $V(\a)$ have the nice property that any pair $V(\a)$ 
and $V(\mathbf{a}')$ is orthogonal except when $\a = -\mathbf{a}'$. To see 
this, take $\eta_{\a}$ in $V(\a)$ and $\eta_{\mathbf{a}'}$ in $V(\mathbf{a}')$. 
The inner form (given by the integral of their wedge product) then transforms 
as 
\begin{equation}
\int_X \eta_{\a} \wedge \eta_{\mathbf{a}'} \rightarrow \int_X \eta_{\a} \wedge 
\eta_{\mathbf{a}'} \prod_i \zeta_i^{a_i + a_i'} \hspace{1cm}  \forall g \in G \, 
.
\end{equation}
However, as the inner form is merely a number which hence must be invariant 
under the action of $G$, it follows that $\a = -\mathbf{a}'$ is a necessary 
condition for the integral to be non-zero.

To see the relation between the forms realized as residues and the eigenspaces 
under the character group, let us use a monomial basis for the polynomials $P$ 
in \eqref{eq:forms_in_PminusX}. For $\mathbf{b} = (b_1, \cdots, b_{5})$, there 
is an associated monomial $\mu_{\mathbf{b}} =  x_0^{b_0} \cdots x_{5}^{b_{5}}$ 
with $\sum_i b_i = \mbox{deg} \, P =  6(4-p)$. To such a monomial, we can 
associate a differential form
\begin{equation}\label{eq:phimonobasis}
\varphi_{\mathbf{a}} =  \frac{\mu_{\mathbf{b}}}{Q(x)^{5-p}}\,\, \Omega_0 \, .
\end{equation}
where $\mathbf{a} =  (1,1,1,1,1,1) + \mathbf{b} $. Under the group action 
\eqref{eq:grouponfermat}, $\varphi_{\mathbf{a}}$ has the simple transformation 
behavior
\begin{equation}
\varphi_{\mathbf{a}}\, \rightarrow \mathbf{a}(g) \varphi_{\mathbf{a}} 
\end{equation}
which follows from the fact that the Fermat polynomial $Q$ of degree $n$ is 
invariant and $\Omega_0$ transforms as 
\begin{equation}
\Omega_0 \rightarrow \left(\prod_i \zeta_i\right) \, \Omega_0 \, .
\end{equation}
As $\mbox{deg}\, P = k \cdot \deg\, Q - 6 =  6k - 6$, we hence have that 
\begin{equation}
|a| \equiv \frac{1}{6} \sum_i a_i = \frac{1}{6}\left(6 + \deg\, P \right) = k = 
5-p\, .
\end{equation}
We can furthermore associate $\mu_{\mathbf{b}}$ with a 
generator of $\mathbb{C} [x_0,\cdots, x_{5}]_{|\mathbf{b}|}/{\langle \partial_i 
Q \rangle}$, as long as $b_i < 5$. When this is satisfied we have $a_i < 6$, so that we conclude that
\begin{equation}
\varphi_{\mathbf{a}} \in H^{p,4-p}(X)\, . 
\end{equation}

This recovers the following result of 
\cite{ASENS_1969_4_2_4_583_0,ogus_crystal,Ran1980,shioda_hodge_conj_fermat_79}, 
which can be phrased in the present context as follows: Let $A^*$ be the subset 
of the character group for which all of the $a_i \neq 0$. Then (Theorem 1 of 
\cite{shioda_hodge_conj_fermat_79}):
\begin{itemize}
 \item[a)] dim$_{\C} V(\a)$ = 1 if and only if $\a \in A^*$; dim$_{\C}V(\a)$ = 0 
otherwise.
 \item[b)] The Hodge type of $V(\a)$ is given by  
 \begin{equation}\label{2eq:hodgetypeeigencycle}
 (p,q) = (5 - |a|,|a|-1) 
 \end{equation}
and the canonical representative with $1 \leq a_i \leq 5$ should be chosen for 
each $a_i$ in the above formula. Note that $|a|$ is always an integer as $\sum 
a_i = 0 $ modulo $6$. Together with a), the above implies that for $\eta_\a \in 
V(\a)$, $\bar{\eta}_{\a}$ is proportional to $\eta_{-\a}$ .
\end{itemize}
It is not hard to use this description to simply enumerate primitive forms by 
counting appropriate tuples $\mathbf{a}$, one finds
\begin{equation}
\begin{array}{c|c}
|\a| & \# \,\, \mbox{elements in } A^* \\
\hline
1 & 1 \\
2 & 426\\
3 & 1751\\
4 & 426\\
5 & 1
\end{array} 
\end{equation}

\subsection{Eigencycles and algebraic cycles}\label{sect:eigencycles}

Elements of $V(\a)$ can not only be realized in terms of $\varphi_{\a}$ but 
also by forming appropriate linear combinations of algebraic cycles, which 
links the two descriptions of elements of $H^{2,2}(X)$. Furthermore, this 
allows 
us to work out the span of the algebraic cycles. Finally, 
finding representatives for all $V(\a)$ with $|\a|=3$ in terms of algebraic 
cycles proves the Hodge conjecture (over $\mathbb{Q}$) for the Fermat sextic, 
see \cite{Ran1980,shioda_hodge_conj_fermat_79,Aoki1983,aoki1987_alg_cycles} for 
more details and generalizations to other Fermat varieties. 

The description of forms $\eta_{\a} \in V(\a)$ in terms of algebraic cycles 
works by putting restrictions on the tuples $\a \in A^*$. We call an element 
$\a \in A^*$ $n$-decomposable if the elements of $\a$ can be decomposed into 
pairs such that maximally $n$ of them satisfy
\begin{equation}
 a_i + a_j = 0 
\end{equation}
modulo $6$. For the Fermat sextic, $\a \in A^*$ with $|\a|=3$, so that 
it corresponds to a $(2,2)$-form, can be $3$-decomposable, $1$-decomposable, or 
indecomposable\footnote{As $\sum a_i =0$ mod $6$, $2$-decomposable implies 
$3$-decomposable.}. Using a computer makes it easy to enumerate them, 
the resulting numbers and their general forms (up to permutations and taking 
the inverse) are given below 
\begin{equation}
\begin{aligned} \mbox{type}&& \mbox{number} && \mbox{standard form}    \\ 
3-\mbox{decomposable}  && 1001 && (r,6-r,s,6-s,t,6-t)\\
1-\mbox{decomposable}  && 720 && (t,6-t,1,3,4,4)\\
\mbox{indecomposable} && 30 && (1,1,4,4,4,4) 
\end{aligned}
\end{equation}
As they should, these sum up to the total $1751$ primitive classes in 
$H^{2,2}(X)$.

Let us consider $3$-decomposable elements of $A^*$. We can write a general 
3-decomposable $\a$ as
\begin{equation}\label{2Sext4fold3Decomsablea}
 a_{\sigma(0)} + a_{\sigma(1)} = 0 \hspace{1cm} a_{\sigma(2)} + a_{\sigma(3)} =0 
\hspace{1cm} a_{\sigma(4)} + a_{\sigma(5)} =0 \, ,
\end{equation}
for some permutation $\sigma$. The corresponding element of $V(\a)$ is
\begin{equation}
\eta_\a = \sum_{\ell_0\ell_1\ell_2} e^{- \frac{i\pi}{3} (a_{\sigma(1)} \ell_0 + 
a_{\sigma(3)} \ell_1 + a_{\sigma(5)} \ell_2)}  
C_{\mathbf{\sigma}}^{\,\boldsymbol{\ell}} \, ,
\end{equation}
where $C_{\mathbf{\sigma}}^{\,\boldsymbol{\ell}}$ are the linear algebraic 
cycles defined in \eqref{eq:cjkmsexticv2}. Using the transformation 
behavior of the $C_{\mathbf{\sigma}}^{\,\boldsymbol{\ell}}$ it is not hard to 
see that it is crucial for the defining relation \eqref{eq:defeigencycle} of 
eigencycles to hold that we are only talking about 3-decomposable $\bf{a}$ 
here. 

This result can be immediately used to constrain the possible intersections 
between the $C_{\mathbf{\sigma}}^{\,\boldsymbol{\ell}}$ and the residues of the 
forms $\varphi_{\a}$, and we shall see how these are in fact fixed up to 
normalization later. A second application concerns the linear relations between 
the $C_{\mathbf{\sigma}}^{\,\boldsymbol{\ell}}$. We have already seen that they 
obey the `sum rule' \eqref{eq:sumrulelincycles} using elementary methods. This 
is insufficient to work out the dimensionality of span of all of the 
$C_{\mathbf{\sigma}}^{\,\boldsymbol{\ell}}$, however. The above proves that its 
dimension is 1001 and shows how further linear relations arise: whenever $\a$ 
is 
3-decomposable in more than one way, we can write down $\eta_\a$ in two 
independent ways in terms of the $C_{\mathbf{\sigma}}^{\,\boldsymbol{\ell}}$ 
using different permutations. As $V(\a)$ is complex one-dimensional, this 
implies that the two expressions must be proportional.

Following the formulae in \cite{aoki1987_alg_cycles}, it is possible to 
write down similar expressions for eigencycles for 1- or in-decomposable $\a$  
using the non-linear algebraic cycles \eqref{eq:Ckjmell} and 
\eqref{eq:Ck1k2k4k5n}.

\subsection{Complex structure moduli}\label{sect:csmoduliresidues}

Having explained how to capture the middle cohomology in terms of residues and 
sketched the relationship to algebraic cycles, let us now discuss complex 
structure deformations in this language. We focus again on the Fermat sextic 
hypersurface in $\P^5$ and consider deforming away from the Fermat locus. We may 
parametrize a general deformation as
\begin{equation}\label{eq:defsextic}
Q(x;s) = \sum_i x_i^{6} + \sum_{\mathbf{b}_I} s_I \,\,\mu_{\mathbf{b}_I} \, 
\end{equation}
for complex parameters $s_I$ and monomials 
\begin{equation}
\mu_{\mathbf{b}_I} =  x_0^{(b_I)_0} \cdots x_{5}^{(b_I)_{5}}
\end{equation}
which are such that $|\mathbf{b}_I|=1$ and $(b_{I})_i < 5$. 

This corresponds to complex structure deformations, which may be represented 
by deformations of the holomorpic top-form $\Omega$, which in turn can be 
written as a residue 
\begin{equation}\label{varphi064fold}
\Omega(s) = \mbox{Res} \left[ \frac{\Omega_0}{Q(x;s)} \right] = 
\mbox{Res} \left[\varphi_{\boldsymbol{1}}\right] \, 
\end{equation}
throughout the moduli space. Setting $s=0$ in the above, we recover the 
holomorphic top-form at the Fermat locus.

The variation of Hodge structure is described by choosing a topological 
basis $\gamma_k$ of $H^4(X)$ and studying the variation of the integrals
\begin{equation}
\int_{\gamma_k} \varphi = \int_{\gamma_k} \mbox{Res} \left[ 
\frac{P}{Q(x;s)^{5-p}}\Omega_0 \right]\, .
\end{equation}
as we vary $Q$. This defines the Hodge bundle and we may locally choose a 
trivialization by identifying the topological cycles $\gamma_k$ in nearby 
sextics. There is a flat connection $\nabla_I$ on this bundle, called the 
Gauss-Manin connection, which acts on residues as
\begin{equation}
 \nabla_I  \varphi =\mbox{Res} \left[\partial_I \frac{P}{Q(x;s)^{5-p}}\Omega_0 
\right]\, .
\end{equation}
The flatness of this connection simply follows from the commutativity of the 
differential operators. 

An infinitesimal deformation of 
\begin{equation}
\Omega = \mbox{Res} \left[\varphi_{\boldsymbol{1}}\right] = \mbox{Res} 
\left[\frac{1}{Q(x;s)}\Omega_0 \right] 
\end{equation}
at the Fermat point can hence be written as
\begin{equation}
\varphi = \left.\varphi_{\boldsymbol{1}}\right|_{s=0} +  \sum_I s_I  
\left.\partial_{s_I}\varphi_{\boldsymbol{1}}\right|_{s=0} \, .
\end{equation}
Note that 
\begin{equation}
\left.\partial_I\varphi_{\boldsymbol{1}}\right|_{s=0} =  
-\frac{\mu_{\mathbf{b}_I} }{Q(x;0)}\Omega_0 = -\varphi_{\mathbf{a}_I}\, ,
\end{equation}
which gives a $(3,1)$-form upon taking the residue at the Fermat point as 
we have restricted to $b_i < 5$ in \eqref{eq:defsextic}. We hence recover that 
deformations of the complex structure are given by $(3,1)$-forms. 

One could also include terms in the sum in \eqref{eq:defsextic} 
for which $\mu_{\mathbf{b}_I}$ is in the Jacobi ideal of $Q$. Deforming by such 
terms again adds a term to $\Omega$ which is given by the residue of a rational 
form, but now the pole order of this rational form can be reduced to $1$ (see 
Appendix \ref{App:RationalForms}). This implies that the residue does not 
produce a $(3,1)$ form, but a $(4,0)$ form. Such deformations would hence only 
rescale $\Omega$. 

In physics, one is usually interested in the covariant derivative $D_I = 
\nabla_I + \partial_I K$, which by definition maps 
\begin{equation}
 D: H^{p,4-p} \rightarrow H^{p-1,4-p+1} \, .
\end{equation}
When working at the Fermat point and acting on $H^{4,0}$, we have just seen that 
we must have $\left.\partial_I K= 0\right|_{s=0} $ as $\nabla_I$ alone already 
has the property of mapping purely to $H^{3,1}$ when using the basis of 
monomials $\mathbf{b}_I$ with $(\mathbf{b}_I)_i < 5$ to define local coordinates 
on the complex structure moduli space\footnote{Note that this does not imply 
that $\partial_I K= 0$ holds for any choice of coordinates on complex structure 
moduli space, as such coordinate changes can give K\"ahler transformation that 
map $K(s,\bar{s}) \rightarrow K(s,\bar{s}) + f(s) + \bar{f}(\bar{s})$ for a 
holomorphic function $f(s)$.}. In the monomial basis we have chosen, the action 
of covariant derivatives is hence particularly simple. 

This structure becomes slightly more complicated when considering second 
derivatives of $\Omega$
\begin{equation}\label{derivativesPhi}
\left. \partial_{I} \partial_J  \varphi_{\boldsymbol{1}}(s) 
\right|_{s=0} = 
\left. 2\, \frac{\mu_{\mathbf{b}_I+\mathbf{b}_{J}}}{Q^{3}}\Omega_0 
\right|_{s=0} 
= 2 \left. \varphi_{\boldsymbol{1} + \mathbf{b}_I + \mathbf{b}_J} \right|_{s=0} 
\, .
\end{equation}
As long as all components of $\mathbf{b}_I+\mathbf{b}_J$ are smaller than $5$, 
this 
form is of pure type $(2,2)$. Whenever this is not the case, however, 
$\mu_{\mathbf{b}_I+\mathbf{b}_{J}}$ is in the Jacobi ideal of $Q$ and we may 
reduce the pole order leading to a form of degree $(3,1)$. To define a covariant 
derivative acting on $(3,1)$-forms, we need to subtract the $(3,1)$-pieces of 
the derivatives. This means we need to set the derivative to zero whenever it 
produces a form for which $\mu_{\mathbf{b}_I+\mathbf{b}_J}$ is in the Jacobi 
ideal of $Q$. 

In summary, the covariant derivative acts on forms as
\begin{equation}
D_I: \mbox{Res}\left[ \varphi_{\mathbf{a}}\right] \rightarrow  \mbox{Res}\left[ 
\varphi_{\mathbf{a}+ \mathbf{b}_I} \right]
\end{equation}
as long as $(\mathbf{a}+ \mathbf{b}_I)_i < 6$ for all $i$ and it sends them to 
zero otherwise. 

We need to evaluate the rank of the matrix \eqref{eq:matrixfluxrank} in order to 
find the (co)-dimension of the Hodge locus and hence the number of stabilized 
moduli. From the above it follows that it can be simply written as   
\begin{equation}\label{eq:matricHodgelocusfinal}
G_{IJ} = D_I D_J \int_X G_4 \wedge \Omega  = 2 \int_X G_4 \wedge \mbox{Res} 
\left[ \frac{\mu_{\mathbf{b}_I + \mathbf{b}_J}}{Q^3}\Omega_0\right] 
\end{equation}
evaluated at the Fermat point. Note that we might as well have written partial 
derivatives as the integral automatically picks out the $(2,2)$ piece of the 
derivatives acting on $\Omega$. In a similar vein, any term for which one of the 
$(\mathbf{b}_I+\mathbf{b}_J)_i \geq 5$ vanishes. See  
\cite{griffiths_carlson_83,voisin_2003,voisin2010hodge,2014arXiv1411.1766M,
movasati_book} for an in-depth discussion of the above result.

\subsection{Period integrals}

In order to evaluate the integral in \eqref{eq:matricHodgelocusfinal}, we need 
to know the period integrals of algebraic cycles, i.e. for an algebraic cycle 
$C$, we need to know
\begin{equation}\label{eq:generalperiodalg}
\int_C  \mbox{Res} \left[ \frac{\mu_{\mathbf{b}_I + 
\mathbf{b}_J}}{Q^3}\Omega_0\right] \, .
\end{equation}

The periods of forms such as \eqref{derivativesPhi} over the linear algebraic 
cycles $C_{\mathbf{\sigma}}^{\,\boldsymbol{\ell}}$  have been computed in 
\cite{movasati_loyola_17,2018arXiv181203964V} using results of 
\cite{griffiths_carlson_83}. The upshot is that for $|\mathbf{b}| = 2$, we have 
that
\begin{equation}\label{eq:movisato_loyola_p2intersections}
\frac{1}{(2 \pi i)^2} \int_{C_\sigma^{\,\boldsymbol{\ell}}} 
\frac{\mu_{\mathbf{b}}}{Q^3}\Omega_0 = 
\left\{
\begin{array}{ll}
\frac{{\rm sgn}(\sigma)}{6^3 2!} \,e^{ \frac{i\pi}{6} \left(\sum_{e=0}^2 
(b_{\sigma(2e)}+1)(2\ell_{e}+1) \right)} & \mbox{if }\, 
b_{\sigma(2e-2)}+b_{\sigma(2e-1)} = 4 \\ \\
0 & \mbox{otherwise.}\\
\end{array}\right.
\end{equation}
Up to the overall normalization, this can also be derived by using the 
automorphism group $(\mathbb{Z}/6 \mathbb{Z})^6/(\mathbb{Z}/6 \mathbb{Z}) 
\rtimes \mathcal{S}_6$ of the sextic. One must have that
\begin{equation}\label{automorphOfX}
\sigma \circ g \left( \int_{C_\sigma^{\,\boldsymbol{\ell}}} 
\frac{\mu_{\mathbf{b}}}{Q^3}\Omega_0 \right)
=\int_{C_\sigma^{\,\boldsymbol{\ell}}} \frac{\mu_{\mathbf{b}}}{Q^3}\Omega_0 
\qquad\qquad g \in (\mathbb{Z}/6 \mathbb{Z})^6/(\mathbb{Z}/6 
\mathbb{Z})\,,\,\,\,\sigma \in \mathcal{S}_6\:,
\end{equation}
with 
$\sigma \circ g \left( \int_{C_\sigma^{\,\boldsymbol{\ell}}} 
\frac{\mu_{\mathbf{b}}}{Q^3}\Omega_0 \right)\equiv 
\int_{\left(C_\sigma^{\,\boldsymbol{\ell}}\right)'} 
\frac{\mu'_{\mathbf{b}}}{Q^3}\Omega'_0$, where the prime quantities are the 
ones 
transformed by $g$ and $\sigma$. 

Let us first consider permutations. After acting with any permutation $\sigma$, 
we may simply relabel the coordinates $x_i$ in the rhs of  \eqref{automorphOfX} 
to undo the permutation again. This produces the same expression we started 
from, except for $\Omega_0$, which produces a sign ${\rm sgn}(\sigma)$ as it is 
completely antisymmetric in the $x_i$. This explains the corresponding factor in 
\eqref{eq:movisato_loyola_p2intersections}. 

Now consider the action by $g \in G = (\mathbb{Z}/6 \mathbb{Z})^6/(\mathbb{Z}/6 
\mathbb{Z})$. This will both act on the differential form under the integral, as 
well as the cycle $C_{\mathbf{\sigma}}^{\,\boldsymbol{\ell}}$. We can write
\begin{equation}
g\left( \int_{C_\sigma^{\,\boldsymbol{\ell}}} 
\frac{\mu_{\mathbf{b}}}{Q^3}\Omega_0 \right)
=\int_{\left(C_\sigma^{\,\boldsymbol{\ell}}\right)'} 
\frac{\mu'_{\mathbf{b}}}{Q^3}\Omega_0 
=\a(g) \int_{C_\sigma^{\,\boldsymbol{\ell}'}} 
\frac{\mu_{\mathbf{b}}}{Q^3}\Omega_0 
\end{equation}
for some $\boldsymbol{\ell}'$, where $\a$ is the element of the character group 
associated with $\a= \mathbf{b} + (1,1,1,1,1,1)$.  To check if this makes 
\eqref{automorphOfX} consistent with
\eqref{eq:movisato_loyola_p2intersections}, it is enough to consider one of the 
generators of $G$, all other cases can be found by analogous computations or 
repeated application of this action. As we have already understood the action of 
permutations, let us furthermore choose $\sigma$ as the trivial permutation 
$\sigma = \id$ and investigate integrals over the cycles 
$C_{\id}^{\,\boldsymbol{\delta}}$. Consider the map $\zeta_0:x_0 \rightarrow 
e^{i \pi /3} x_0$, which generates one of the $(\mathbb{Z}/6 \mathbb{Z}) \subset 
G$. $\zeta_0$ maps $C_\sigma^{\,\boldsymbol{\ell}}$ to 
$C_\sigma^{\,\boldsymbol{\ell}'}$ where $\delta_0' = \delta_0 -1 $. We can hence 
write
\begin{eqnarray}
a(\zeta_0)\int_{C_{\id}^{\,\boldsymbol{\delta}'}} 
\frac{\mu_{\mathbf{b}}}{Q^3}\Omega_0  &=& e^{\frac{i\pi}{3}a_0} 
\int_{C_{\id}^{\,\boldsymbol{\delta}'}} \frac{\mu_{\mathbf{b}}}{Q^3}\Omega_0 
\nonumber\\
&=& e^{\frac{i\pi}{3}(b_0+1)}
 \frac{1}{6^3 2!} e^{\frac{i \pi}{6} \left[(b_{0}+1)(2 \ell_0'+1)+(b_{2}+1)(2 
\ell_1'+1)+(b_{4}+1)(2 \ell_2'+1)\right] }   \nonumber\\
 &=& e^{\frac{i\pi}{3}(b_0+1)}
 \frac{1}{6^3 2!} e^{\frac{i \pi}{6} \left[(b_{0}+1)(2 \ell_0-1)+(b_{2}+1)(2 
\ell_1+1)+(b_{4}+1)(2 \ell_2+1)\right] }   \nonumber\\
  &=& 
 \frac{1}{6^3 2!} e^{\frac{i \pi}{6} \left[(b_{0}+1)(2 \ell_0+1)+(b_{2}+1)(2 
\ell_1+1)+(b_{4}+1)(2 \ell_2+1)\right] }   \nonumber\\
&=& \int_{C_{\id}^{\,\boldsymbol{\delta}}} \frac{\mu_{\mathbf{b}}}{Q^3}\Omega_0
\end{eqnarray}
that is exactly what \eqref{automorphOfX} says.

We then need to know the integral of $\varphi_\a$ on a  single cycle 
$C_\sigma^{\boldsymbol{\ell}}$ to compute the integrals of $\varphi_\a$ over all 
the cycles $C_\sigma^{\boldsymbol{\ell}'}$ in the same orbit. One simply uses
\begin{equation}
 \int_{C_\sigma^{\boldsymbol{\ell}'}} \varphi_{\bf a} = \a(g)^{-1}  
\int_{C_\sigma^{\boldsymbol{\ell}}} \varphi_{\bf a} \,.
\end{equation}
that is derived by \eqref{automorphOfX}.
This shows that in fact all relative coefficients of 
\eqref{eq:movisato_loyola_p2intersections} are fixed by $G\rtimes 
\mathcal{S}_6$, as it acts transitively on the 
$C_\sigma^{\,\boldsymbol{\ell}}$. It is not true, however, that $G\rtimes 
\mathcal{S}_6$ acts transitively on a basis of algebraic cycles for $H^{2,2}(X) 
\cap H^4(X,\mathbb{Q})$. If we want to study periods of such a basis up to a 
global normalization, we hence need more than the relative factors between 
periods of the 
$C_\sigma^{\,\boldsymbol{\ell}}$.

Note that the condition  $b_{\sigma(2e-2)}+b_{\sigma(2e-1)} = 4$ for all 
$e\in\{0,1,2\}$ implies that the intersections of 
$C_\sigma^{\,\boldsymbol{\ell}}$ with the eigenspace $V(\a)$ is non-zero only if 
$\a$ is 3-decomposable. This is not unexpected, as we have seen, $V(\a)$ for 
$\a$ 3-decomposable can be constructed from the  
$C_\sigma^{\,\boldsymbol{\ell}}$, whereas eigenspaces for $\a$ 1-decomposable or 
indecomposable are constructed from other algebraic cycles. The vanishing 
statement hence strengthens the observation that $V(\a)$ and $V(\a')$ are 
orthogonal except $\a' = \bar{\a}$. This can also be seen directly as follows.
Let us consider the case where $\sigma$ is the trivial permutation. The action 
of $\zeta_0^{k} \zeta_1^k$ on $C_\sigma^\ell$ is trivial in this case. It 
follows that
\begin{equation}
\int_{C_\sigma^{\,\boldsymbol{\ell}}} \frac{\mu_{\mathbf{b}}}{Q^3}\Omega_0 
=
\zeta_0^{k} \zeta_1^k \int_{C_\sigma^{\,\boldsymbol{\ell}}} 
\frac{\mu_{\mathbf{b}}}{Q^3}\Omega_0 
= e^{\frac{i\pi}{3} \cdot k\left(b_0+b_1+2\right)}
\int_{C_\sigma^{\,\boldsymbol{\ell}}}\frac{\mu_{\mathbf{b}}}{Q^3}\Omega_0 
\end{equation}
so that the integral can only be non-zero when $a_0 + a_1 = 0 \mod 6$. We can 
make the same argument for the other two pairs $x_2,x_3$ and $x_4,x_5$. The same 
argument applies (with different pairings) for other permutations, and implies 
that $\a = \mathbf{b}+1$ must be 3-decomposable for the integral to be 
non-zero. 

\

The above can be generalized to arbitrary algebraic cycles of complete 
intersection type \cite{2018arXiv181203964V}, i.e. cycles of the type 
$f_0=f_1=f_2 = 0$ inside a hypersurface 
\eqref{eq:factorization_for_cialgcycle}. The result is
\begin{equation}\label{eq:movisato_loyola_generalintersections}
\frac{1}{(2 \pi i)^2}\int_Z \frac{\mu_{\mathbf{b}}}{Q^3}\Omega_0 = c \cdot 
\frac{5^6}{2}\, ,
\end{equation}
where $c$ is the unique number which satisfies 
\begin{equation}\label{eq:def_c_loyola}
\mu_{\mathbf{b}} \det(\partial_i H_j ) =  c\,\, \det\left({\rm Hess}(Q)\right) 
\hspace{0.5cm} \mod \langle \partial_i Q \rangle \, ,
\end{equation}
the vector $H$ is given by $H = \left(f_0,g_0,f_1,g_1,f_2,g_2\right)$ and Hess 
denotes the Hessian matrix. For the linear cycles $C^\ell_\sigma$, this 
reproduces the normalization of \eqref{eq:movisato_loyola_p2intersections} from 
the general formula \eqref{eq:movisato_loyola_generalintersections}.

\section{Algebraic Fluxes and Stabilization of Complex Structure Moduli} 
\label{sect:fluxstab}

With the tools we have collected in the previous section, we are now ready to 
directly address how algebraic cycles can be used as fluxes and how many moduli 
they stabilize. All one needs to do after defining a flux which is 
appropriately 
quantized and primitive, is to evaluate the period integrals 
\eqref{eq:generalperiodalg} needed to compute the rank of the $426 \times 426$ 
matrix $G_{IJ}$ \eqref{eq:matricHodgelocusfinal}.

\subsection{One linear algebraic cycle}

As a first example, let us revisit the case of using a single linear algebraic 
cycle $C_{\mathbf{\sigma}}^{\,\boldsymbol{\ell}}$ \eqref{eq:cjkmsexticv2} as a 
$G_4$ flux.\footnote{The same issues as discussed in footnote \ref{footnote3} 
apply.} Using the period integrals \eqref{eq:movisato_loyola_p2intersections} we 
find
\begin{equation}
\mbox{rk} G_{IJ} \left( C_{\mathbf{\sigma}}^{\,\boldsymbol{\ell}} \right) = 19 
\, , 
\end{equation}
which is precisely the same number we obtained by analyzing obstructed 
polynomial deformations. 

In fact, this is the lowest possible value $G_{IJ}$ can have for any algebraic 
cycle. This is not surprising as linear algebraic cycles are the simplest type 
that can exist for the Fermat sextic (see  Proposition 7 `Olympiad problem' of 
\cite{2014arXiv1411.1766M}).

\subsection{A sum of two linear algebraic cycles}

With the material we have collected, it is straightforward to work out what 
happens when we add two different linear algebraic cycles 
$C_{\mathbf{\sigma}}^{\,\boldsymbol{\ell}} + 
C_{\sigma'}^{\,\boldsymbol{\ell}'}$. 
A direct computation (in combination with the automorphism group) shows that 
the rank of $G_{IJ}$ only depends on the mutual intersection between the two, 
and we can make the following table
\begin{equation}
\begin{array}{c|c}
C_{\mathbf{\sigma}}^{\,\boldsymbol{\ell}} \cdot 
C_{\mathbf{\sigma}'}^{\,\boldsymbol{\ell}'} & \mbox{rk} \,\, G_{IJ}\left( 
C_{\mathbf{\sigma}}^{\,\boldsymbol{\ell}} + 
C_{\mathbf{\sigma}'}^{\,\boldsymbol{\ell}'} \right)\\ 
\hline
21&  19\\
-4&  32\\
1& 38 \\
0 & 38
\end{array}\, . 
\end{equation}
The first row corresponds to the case 
$C_{\mathbf{\sigma}}^{\,\boldsymbol{\ell}}= 
C_{\mathbf{\sigma}'}^{\,\boldsymbol{\ell}'}$. As the number of flat directions 
for a linear combination of two cycles is at least equal to the number of flat 
directions common to both of them, the rank of the matrix $G_{IJ}$ must be 
subadditive:
\begin{equation}\label{eq:subadditivity}
\mbox{rk} \,\,G_{IJ} \left( C_{\mathbf{\sigma}}^{\,\boldsymbol{\ell}} \right)
+ 
\mbox{rk} \,\, G_{IJ} \left( C_{\mathbf{\sigma}'}^{\,\boldsymbol{\ell}'} \right)
\geq \mbox{rk} \,\, G_{IJ} \left( C_{\mathbf{\sigma}}^{\,\boldsymbol{\ell}} + 
C_{\mathbf{\sigma}'}^{\,\boldsymbol{\ell}'} \right)\, ,
\end{equation}
which is indeed the case for the numbers we find. 

\subsection{Fluxes respecting group 
actions}\label{sect:flux_example_GP_symmetry}

The sextic moduli space has the symmetry group $G=\mu^{6}/\mu$ for $\mu = \Z/6\Z$, that we discussed in Section~\ref{sect:group_on_residues}. 
Consider the Greene-Plesser subgroup $G_{PG} = (\Z/6\Z)^4$ \cite{Greene:1990ud}. 
It 
is generated by $\alpha_i^6=1$ for $i = 1,2,3,4$ with action
\begin{equation}
(x_0,x_1,x_2,x_3,x_4,x_5) \rightarrow  
((\alpha_1\alpha_2\alpha_3\alpha_4)^{-1} x_0,\alpha_1 x_1,\alpha_2 x_2,\alpha_3 
x_3,\alpha_4 x_4,x_5)
\end{equation}
on the homogeneous coordinates of $X_6$. Famously, only a single complex 
structure deformation, corresponding to the monomial $\prod_i x_i$, i.e. 
$\mathbf{b} = (1,1,1,1,1,1)$, is symmetric under the action of this group, 
while all others are projected out. The obvious way to construct a flux that is 
even under the action of $G_{PG}$ is to start with the orbit of any of 
the linear cycles $C_{\mathbf{\sigma}}^{\,\boldsymbol{\ell}}$ under $G_{PG}$. 
It turns out that this is not the minimal choice and one can repeatedly use the 
sum rule \eqref{eq:sumrulelincycles} to show that
\begin{equation}
\sum_{g \in G_{GP}} g(C_{\mathbf{\sigma}}^{\,\boldsymbol{\ell}}) =  4\, 
C_{eee}
\end{equation}
where
\begin{equation}
C_{eee} = \sum_{\ell_0,\ell_1,\ell_2 \in (0,2,4)} 
C_{\mathbf{\sigma}}^{\,\boldsymbol{\ell}} \, .
\end{equation}
Using \eqref{eq:sumrulelincycles} and the intersection numbers 
\eqref{eq:int_numbers_lin_cycles} one can also show directly that $C_{eee}$ is 
even under $G_{PG}$ and that $C_{eee} \cdot C_{eee} = 3^5$. As there are $3^3$ 
terms in the sum, we have $C_{eee} \cdot H^2 = 27$. A symmetric flux that is 
primitive and properly quantized is 
\begin{equation}
 G_4^{\rm sym} = C_{eee} - 9/2 H^2 \,,
\end{equation}
and the induced tadpole is hence $243/4$, which is well within the allowed 
range.

Directly evaluating $G_{IJ}(C_{eee})$ using 
\eqref{eq:movisato_loyola_p2intersections} we find that 
indeed
\begin{equation}
\mbox{rk} \,\,G_{IJ} \left(  G_4^{\rm sym} \right) = 141 \,.
\end{equation}
In particular, there is a single entry that obstructs the unique deformation 
that is symmetric under $G_{PG}$. Hence the symmetric flux fixes the invariant 
modulus; this corresponds to solving $D_{s^{\rm i}}W=0$ explicitly (see at the 
end of Section~\ref{sect:background} for notation) and finding that the solution 
sits at the Fermat point. Our computation goes further: it is true that the 
Fermat point (that belongs to the fixed point set of $G_{PG}$) is a solution of 
$D_{s^{\rm ni}_a}W=0$ (with $a=1,...,425$), but 
only $140$ out of the $425$ 
non-symmetric deformations under $G_{PG}$ are fixed by $G_4^{\rm sym}$, the 
other $285$ ones are flat directions. 

Finally, one may wonder about using a flux that is symmetric under the entire 
automorphism group $(\mathbb{Z}/6 \mathbb{Z})^6/(\mathbb{Z}/6 \mathbb{Z}) 
\rtimes \mathcal{S}_6$ of the sextic at the Fermat point. As all of the forms 
$\varphi_\a$ with $|a|=3$ have non-trivial transformations already under the 
scaling part, it follows that the matrix $G_{IJ}$ can only contain zeros in 
such a case. The same can be seen by noting that 
$(\mathbb{Z}/6 \mathbb{Z})^6/(\mathbb{Z}/6 \mathbb{Z}) 
\rtimes \mathcal{S}_6$ acts transitively on the 
$C_{\mathbf{\sigma}}^{\,\boldsymbol{\ell}}$. Using the sum rule  
\eqref{eq:sumrulelincycles} one can argue that summing over an orbit results in 
a cycle proportional to $H^2$ (the only invariant cycle), so that $G_{IJ}$ 
vanishes for all $I,J$. This implies that there are no invariant fluxes that 
are primitive for this group.

\subsection{Stabilizing all moduli using linear algebraic cycles}

Let us now see if we can find a flux stabilizing all moduli employing only 
linear algebraic cycles. From the subadditivity \eqref{eq:subadditivity}, it 
follows that we need to consider a linear combination of at least $23$ of the 
$C_{\mathbf{\sigma}}^{\,\boldsymbol{\ell}}$. As we have seen in Sections 
\ref{sect:group_on_residues} and \ref{sect:eigencycles}, only a subspace of 
dimension $1001$ within $H^{2,2}_{\mbox{\rm prim}}(X)$ is spanned by the 
linear algebraic cycles, and this subspace precisely corresponds to 
3-decomposable tuples $\a$. By a simple scan over all possibilities, one can 
find that for every $b_I$ with $|\mathbf{b}_I|=1$, there is a $b_J$ with  
$|\mathbf{b}_J|=1$ such that
\begin{equation}
\a_{IJ} = (1,1,1,1,1,1) + b_I + b_J 
\end{equation}
is 3-decomposable. In other words, linear cycles in principle allow us to 
constrain all complex structure deformations. 

Due to the large number of linear algebraic cycles, there are 
$6^3 \cdot 15  = 3240$ of them, a simple scan is computationally much too 
expensive. Besides randomly sampling choices, the following (semi-)systematic 
method can be used. It is an experimental fact that the inequality 
\eqref{eq:subadditivity} becomes an equality if we consider sums of linear 
algebraic cycles such that all of them are mutually orthogonal, i.e.
\begin{equation}\label{eq:additivity}
\mbox{rk} \,\,G_{IJ} \left(  \sum_{i \in I} C_{\sigma_i}^{\,\boldsymbol{\ell_i}} 
\right) = |I| \cdot 19
\qquad\mbox{if}\qquad
C_{\sigma_i}^{\,\boldsymbol{\ell_i}}  \cdot C_{\sigma_j}^{\,\boldsymbol{\ell_j}} 
= 0  
\end{equation}
for all $i \neq j \in I$. Correspondingly, the maximal size of a set with this 
property is $22$, one possible choice being
\begin{equation}
\begin{aligned}
I_{\mbox{\tiny max}} = \{&\left[\left[0, 0, 0\right], \left[0, 1, 2, 3, 4, 
5\right]\right], \left[\left[0, 0, 0\right], \left[0, 2, 1, 5, 3, 
4\right]\right], \left[\left[0, 0, 1\right], \left[0, 3, 1, 2, 4, 
5\right]\right], \\
&
\left[\left[0, 0, 1\right], \left[0, 5, 1, 3, 2, 4\right]\right],
\left[\left[0, 1, 0\right], \left[0, 4, 1, 2, 3, 5\right]\right],
\left[\left[1, 0, 1\right], \left[0, 1, 2, 5, 3, 4\right]\right],\\
&
\left[\left[1, 0, 2\right], \left[0, 3, 1, 4, 2, 5\right]\right],
\left[\left[1, 1, 1\right], \left[0, 2, 1, 4, 3, 5\right]\right],
\left[\left[1, 1, 4\right], \left[0, 4, 1, 3, 2, 5\right]\right],\\
&
\left[\left[1, 2, 2\right], \left[0, 5, 1, 2, 3, 4\right]\right],
\left[\left[2, 0, 5\right], \left[0, 1, 2, 4, 3, 5\right]\right],
\left[\left[2, 1, 4\right], \left[0, 3, 1, 5, 2, 4\right]\right],\\
&
\left[\left[2, 2, 2\right], \left[0, 2, 1, 4, 3, 5\right]\right],
\left[\left[2, 3, 1\right], \left[0, 4, 1, 3, 2, 5\right]\right],
\left[\left[3, 2, 3\right], \left[0, 1, 2, 4, 3, 5\right]\right],\\
&
\left[\left[3, 2, 5\right], \left[0, 4, 1, 3, 2, 5\right]\right],
\left[\left[3, 3, 2\right], \left[0, 5, 1, 4, 2, 3\right]\right],
\left[\left[3, 4, 3\right], \left[0, 2, 1, 5, 3, 4\right]\right],\\
&
\left[\left[3, 5, 5\right], \left[0, 3, 1, 2, 4, 5\right]\right],
\left[\left[4, 3, 4\right], \left[0, 4, 1, 2, 3, 5\right]\right],
\left[\left[5, 3, 3\right], \left[0, 4, 1, 5, 2, 3\right]\right],\\
&
\left[\left[5, 5, 4\right], \left[0, 2, 1, 5, 3, 4\right]\right]\,\,\}
\end{aligned}\, ,
\end{equation}
where each entry is of the form $[\ell_i,\sigma_i]$. Consistent with 
\eqref{eq:additivity}, a sum of the associated linear algebraic cycles gives a 
matrix $G_{IJ}$ of rank $418$. 

We can use this as a starting point to construct a flux stabilizing all moduli 
by adding a further linear algebraic cycle. We are looking for a $C_e$ such that
for 
\begin{equation}
C =  \sum_{i \in I_{\mbox{\tiny max}}} C_{\sigma_i}^{\,\boldsymbol{\ell_i}} + 
C_e
\end{equation}
the flux 
\begin{equation}
G_4 = C + n H^2
\end{equation}
is primitive and appropriately quantized. Quantization requires $n$ to be 
half-integer, $n=m/2$ for $m$ odd, and primitivity requires
\begin{equation}
3 m  = 22 + C_e \cdot H^2 \, ,
\end{equation}
so that we can choose $C_e = -  C_{\sigma_e}^{\,\boldsymbol{\ell_e}}$, which 
gives $m=7$. The tadpole of such a configuration is given by 
\begin{equation}
N_{D3} = \tfrac12 \left(  \sum_{i \in I_{\mbox{\tiny max}}} 
C_{\sigma_i}^{\,\boldsymbol{\ell_i}} -  C_{\sigma_e}^{\,\boldsymbol{\ell_e}} - 
\frac{7}{2}H^2 \right)^2 = \tfrac12 \left(\left( \sum_{i \in I_{\mbox{\tiny 
max}}} C_{\sigma_i}^{\,\boldsymbol{\ell_i}} -  
C_{\sigma_e}^{\,\boldsymbol{\ell_e}} \right)^2 - 6\frac{7^2}{4}\right)
\end{equation}
which is minimal if we choose $C_{\sigma_e}^{\,\boldsymbol{\ell_e}} \cdot 
\sum_{i \in I_{\mbox{\tiny max}}} C_{\sigma_i}^{\,\boldsymbol{\ell_i}}$ to be as 
large as possible. There is a unique such $C_{\sigma_e}^{\,\boldsymbol{\ell_e}}$ 
which has
$C_{\sigma_e}^{\,\boldsymbol{\ell_e}} \cdot \sum_{i \in I_{\mbox{\tiny max}}} 
C_{\sigma_i}^{\,\boldsymbol{\ell_i}} = 11$, it is given by
\begin{equation}
[\sigma_e,\ell_e] = [[2, 5, 3], [0, 2, 1, 4, 3, 5]] \, .
\end{equation}
Working out $G_{IJ}(C) = G_{IJ}(G_4)$ one finds that it has maximal rank, $426$.
The resulting tadpole is computed to be
\begin{equation}\label{eq:tadpolelincyccombo}
N_{D3} = \tfrac12 G_4 \wedge G_4 = \frac{775}{4} \, . 
\end{equation}
Although such a flux would stabilize all complex structure moduli at the Fermat 
point, it significantly overshoots the available tadpole 
\begin{equation}
\chi(X)/24  = 435/4 \, .
\end{equation}

\subsection{Tadpole issues}

The result of the last section is at the same time encouraging and 
disappointing: while it is not hard to find a primitive flux with proper 
quantization that
stabilizes all moduli, it generates a tadpole that is almost twice the maximal 
allowed value. Scanning over random linear combinations of linear algebraic 
cycles gives many more examples with the same properties. This result can 
already be anticipated from rough estimates.

Consider stabilizing all moduli by a combination of linear algebraic fluxes
\begin{equation}
C = \sum_{i \in I}  f_i C_{\sigma_i}^{\,\boldsymbol{\ell_i}}   
\end{equation}
and let $\sum f_i = 3 m$ for an odd integer $m$. We can then find a primitive 
properly quantized flux by setting
\begin{equation}
G_4 = C - \frac{m}{2} H^2 \, . 
\end{equation}
The induced tadpole is then
\begin{equation}
\tfrac12 G_4^2 = \tfrac12 \left(C^2 - \frac{3}{2} m^2 \right)\,. 
\end{equation}
Ignoring the contribution from $C_{\sigma_i}^{\,\boldsymbol{\ell_i}}   \cdot 
C_{\sigma_j}^{\,\boldsymbol{\ell_j}}$ to $C^2$, and assuming that all $f_i =1$, 
we can write this as
\begin{equation}\label{eq:tadpole_estimate}
\tfrac12 G_4^2 = \tfrac12 \left(63 m- \frac{3}{2} m^2 \right)\,. 
\end{equation}
As every $C_{\mathbf{\sigma}}^{\,\boldsymbol{\ell}}$ stabilizes at most $19$ 
moduli, we need to have $3 m > 426/19$. This roughly reproduces 
\eqref{eq:tadpolelincyccombo} for the minimal choice of $m$.

The negative contribution in \eqref{eq:tadpole_estimate} points at a potential 
way out by letting $m$ become sufficiently large. As we have seen, there are at 
most $22$ mutually orthogonal linear algebraic cycles and we need to let $3 m$ 
be significantly larger to bring the tadpole down sufficiently. It turns out that 
ignoring mutual intersections between the terms in $C$ become increasingly 
unjustified, so that the tadpole contribution of such fluxes is again far too 
large to give a viable model.

Until now, we have completely ignored non-linear algebraic cycles. Performing a 
similar rough estimate gives a comparable result to what we have found for the 
linear algebraic cycles. There, the crucial ratio was that of the number of 
moduli that could be fixed with a single linear algebraic cycle $19$, to the 
square of such a cycle, $21$. These ratios can also be computed for non-linear 
algebraic cycles, the result is that for a complete intersection algebraic cycle 
$C_{f_0 f_1 f_2}$ given by $f_0=f_1=f_2=0$ for homogeneous polynomials with 
degrees $d_i$ \cite{2016arXiv160206607M}
\begin{equation}
 \begin{array}{ccc}
(d_0,d_1,d_2)  & C_{f_0 f_1 f_2}^2 & \mbox{rk} \,\,G_{IJ} \left( C_{f_0 f_1 f_2} \right) \\
\hline 
(1,1,1)  & 21 & 19\\
(1,1,2)  & 34& 32 \\
(1,1,3)  & 39& 37\\
(1,2,2)  & 56& 54\\
(1,2,3)  & 66& 62\\
(1,3,3)  & 81& 71\\
(2,2,2)  & 96& 92\\
(2,2,3)  & 120& 106\\
(2,3,3)  & 162& 122\\
(3,3,3)  & 243& 141
\end{array}\, . 
\end{equation}
As the ratio of these two numbes stays roughly the same, we can anticipate to find similar results using non-linear algebraic cycles. As the maximal allowed tadpole of the flux is $\chi(X)/24 = 435/4 = \tfrac12 C^2$, but we need to stabilize $426$ moduli, the ratio between $C^2$ and $\mbox{rk}\,\,G_{IJ} (C)$ should be roughly $\tfrac12$ rather than the ratio of $\sim 1$ (and larger) observed above.

The above results do not imply that there cannot be a properly quantized and 
primitive flux stabilizing all moduli that also satisfies the tadpole 
constraint. Much more work is needed to make such a claim. What we can say (at 
least for the sextic fourfold we studied), however, is that it is not completely 
straightforward to construct such a flux. 

\section{Moduli Stabilization and Symmetry Actions}\label{sect:symmetry_actions}

In this section we address the problem of moduli stabilization in cases 
with symmetry in some more generality. In particular, we explain why 
fluxes that are invariant under some group actions typically leave some flat 
directions in the effective potential. This will give a conceptual way of 
understanding the result found in Section \ref{sect:flux_example_GP_symmetry}.

Let $X_p$ be a Calabi-Yau fourfold at a point $p$ in its complex structure 
moduli space, and $G$ any subgroup of the automorphism group of $X_p$. Although 
what we are going to say can be put in slightly more general terms, let us 
assume for simplicity that $X_p$ is a hypersurface in a toric 
variety $T$ for which we can represent all forms in the middle cohomology of 
$X_p$ as residues. Let us furthermore assume that $G$ acts by rescaling the 
homogeneous coordinates $x_i$ of the ambient space by roots of unity and 
preserves the holomorphic top form $\left.\Omega_X\right|_p$ at $p$ (as well as 
the K\"ahler form of 
$X_p$). 

We can write a family in the vicinity of $p$ as
\begin{equation}
X: \,\, Q = Q_0 + \sum_N s_N \mu_N + \sum_\Phi t_\Phi \nu_\Phi = 0
\end{equation}
where $X_p$ is given by $\mathbf{s} = \mathbf{t} = 0$, the monomials 
$\mu_N$ are invariant under the action of $G$, and the monomials 
$\nu_\Phi$ are not, but transform as
\begin{equation}
 \nu_\Phi \rightarrow \alpha_\Phi(g)  \nu_\Phi\, \hspace{.5cm} \mbox{(no 
summation)}.
\end{equation}

By assumption we can write
\begin{equation}
\left. \Omega_X \right|_p= \mbox{Res} \left[ \frac{1}{Q_0} \Omega_T\right] \, ,
\end{equation}
for some fixed holomorphic form $\Omega_T$ on the ambient space $T$. As both 
$\Omega_X$ at $p$ and $Q_0$ are invariant under the action of $G$, it follows 
that $\Omega_T$ must be preserved by $G$ as well.

Let us now consider switching on a flux $G_4$ which is invariant under the 
action of $G$. This implies that the GVW superpotential \eqref{GVWsup} is 
invariant under $G$ as well. The F-terms in the non-invariant 
directions $t_\Phi$ at $p$ transform as
\begin{equation}
F_\Phi = \int_{X_p} G_4 \wedge \left.  D_\Phi \Omega_X \right|_p= - \int_{X_p} 
G_4 \wedge  \mbox{Res} \left[ \frac{\nu_\Phi}{Q_0^2} \Omega_T\right] \, 
\rightarrow \alpha_\Phi(g) F_\Phi .
\end{equation}
However, the above integral at $p$ simply yields a number that cannot change 
under any automorphism, so that it follows that $F_\Phi = 0$ for all $\Phi$. 
This argument was used in 
\cite{Giryavets:2003vd,Denef:2004dm,Louis:2012nb,Cicoli:2013cha} to argue that\footnote{
These papers deal with Calabi-Yau threefolds. However their arguments and their 
conclusions directly apply to fourfolds as well.} 
the F-term equations in the non-invariant directions are automatically 
satisfied and one only needs to take care of the invariant directions. 

As we have seen in the example in Section \ref{sect:flux_example_GP_symmetry}, 
this does not imply that there are no flat directions along the non-invariant 
directions $t_\Phi$ in complex structure moduli space. We can repeat a similar 
argument as above for the matrix $G_{IJ}$ to see why. Let us first consider the 
mixed terms $G_{N\Phi}$ between invariant and non-invariant directions. 
They transform as
\begin{equation}
G_{N\Phi} =  \int_{X_p} G_4 
\wedge  \mbox{Res} \left[ \frac{\mu_n \nu_j}{Q_0^3} \Omega_T\right] 
\rightarrow \alpha_\Phi(g) \, G_{N\Phi} \, .
\end{equation}
As these have a non-trivial scaling they must vanish, so that $G_{IJ}$ is 
block-diagonal among invariant and non-invariant directions. For the matrix 
elements between two non-invariant directions we find
\begin{equation}
G_{\Phi \Xi} =  \int_{X_p} G_4 
\wedge  \mbox{Res} \left[ \frac{\nu_\Phi \nu_\Xi}{Q_0^3} \Omega_T\right] 
\rightarrow \alpha_\Phi(g) \alpha_\Xi(g) G_{\Phi\Xi}   \, .
\end{equation}
These can hence only be non-vanishing if $\alpha_\Phi(g) 
\alpha_\Xi(g)=1$ for all $g \in G$. This is a strong condition, and for many 
$\nu_\Phi$ there is no $\nu_\Xi$ such that it holds, which implies that both 
$t_\Phi$ and $t_\Xi$ are flat directions. 

Let us now come back to the example discussed in Section 
\eqref{sect:flux_example_GP_symmetry}, where $X_p$ is the sextic fourfold at the 
Fermat point and $G$ is the Greene-Plesser group $G_{GP}$. In this case there 
is only a single invariant monomial $\mu_N = \prod_i x_i$. All other monomials 
$\nu_\Phi$ correspond to non-invariant directions. In order to have a non-zero 
matrix elements $G_{\Phi\Xi}$ we need that 
\begin{equation}
\nu_\Phi \nu_\Xi = \prod_i x_i^2 \, ,
\end{equation}
as this is the only monomial of appropriate degree that is invariant under 
$G_{GP}$. 
The only non-invariant complex structure deformations that have non-zero 
elements in $G_{IJ}$ hence correspond to pairs of tuples $\mathbf{b}_\Phi$ and 
$\mathbf{b}_\Xi$ different from $(1,1,1,1,1,1)$ with $\sum_i 
(\mathbf{b}_\Phi)_i = 6$ and $\sum_i (\mathbf{b}_\Xi)_i = 6$ such that 
\begin{equation}
 \mathbf{b}_\Phi  + \mathbf{b}_\Xi = (2,2,2,2,2,2)\, .
\end{equation}
It turns out that there are precisely $70$ such pairs, so that the rank of 
$G_{IJ}$ can be at most $141$ for any flux that is symmetric under $G_{GP}$, 
i.e. there are at least $285$ flat directions in this case. For the example we 
have chosen in Section \ref{sect:flux_example_GP_symmetry}, this is precisely 
what was found by explicitly evaluating $G_{IJ}$.

\section{Conclusions and Future Directions}

In this work we have begun to explore how to use algebraic cycles as fluxes on 
Calabi-Yau fourfolds. We have reviewed methods which allow us to compute the 
number of stabilized moduli and the induced tadpole for fluxes proportional to 
linear combinations of algebraic cycles. We have analyzed in detail the sextic 
fourfold. We have found fluxes that stabilize all the complex structure moduli 
at a specific point in the complex structure moduli space, without the need of 
dealing with Picard-Fuchs equations of (very) high rank. What is striking about 
this analysis is that it appears very hard to find a flux that satisfies all 
consistency constraints and stabilizes all of the complex structure moduli.
In particular, in the example we have considered, we have noticed tension 
between  tadpole cancellation and the desire to stabilize all complex structure 
moduli.

The above is far from a complete analysis, and there are several crucial points 
that need to be addressed for a complete picture. First of all, it is in 
principle straightforward (but tedious) to work out $\mbox{rk}\,\ 
G_{IJ}(C_\Sigma)$, $C_\Sigma^2$, and $H \cdot C_\Sigma$ for any linear 
combination of algebraic cycles $C_\Sigma$. Having access to all fluxes defined 
via algebraic cycles is not sufficient, as the integral Hodge conjecture for the 
Fermat sextic is presently unanswered. It has been shown to be correct, however, 
for the quartic and quintic Fermat fourfolds in \cite{ALJOVIN2019177}, and it 
is possible to extend their methods to the sextic. With a proof of the integral 
Hodge conjecture for the Fermat sextic, it is then possible to compute 
$\mbox{rk}\,\ G_{IJ}(G_4)$ for all fluxes satisfying the tadpole constraint. 
This naively seems like a task that is computationally too demanding to be 
undertaken, but a clever exploitation of the large automorphism group of 
the sextic might make it feasible. We intend to attack this problem in future 
work. 

Thinking even further ahead, it is highly desirable to extend the 
methods we have reviewed to other points in the moduli space of the sextic with 
maximal $H^{2,2}(\Z) \cap H^{4}(X,\Z)$, and even to other Calabi-Yau fourfolds. 
In particular, it would be exciting to find criteria which can distinguish 
which points in the moduli space can and which cannot be stabilized using 
fluxes that satisfy the tadpole constraint. Such criteria would have 
far reaching implications for the existence and structure of the string 
landscape.

Finally, let us note that the Fermat sextic we considered in this paper 
is known to be modular \cite{Schimmrigk:2008mp}. In recent work, it was 
conjectured that in fact all flux vacua correspond to modular varieties 
\cite{Kachru:2020sio}, although the converse to this statement is not true 
\cite{Schimmrigk:2020dfl}. A similar statement holds for attractive $K3$ surfaces 
\cite{livne_K3}, which appear in the study of flux vacua on the 'toy' fourfold 
$K3 \times K3$ \cite{Aspinwall:2005ad} (see also \cite{Braun:2014ola}), 
as well as the closely related attractor points on Calabi-Yau 
manifolds \cite{Candelas:2019llw}. At its core, modularity is concerned with 
Galois representations, which are again related to algebraic cycles according to 
the Tate conjecture. It should be fascinating to explore this relationship 
further.

\section*{Acknowledgements}

We wish to thank Hossein Movasati for discussion and explanations. The work of 
R.V.~is partially supported by ``Fondo per la Ricerca di Ateneo - FRA 2018'' 
(UniTS) and by INFN Iniziativa Specifica ST\&FI.

\appendix

\section{Rational Forms, Residues, and Cohomology of Hypersurfaces}\label{App:RationalForms}

In this section we review some classic material concerning rational 
differential forms on $\P^n$, i.e. forms with poles, and their residues. This 
is based on \cite{griffithsI,griffithsII}, a beautiful exposition of which can 
be found in \cite{cox1999mirror,Candelas:2000fq,Doran:2007jw,movasati_book}.

The basic idea of residues of forms is to extend the residue formula
\begin{equation}
\frac{1}{2 \pi i}\int_\gamma \frac{dz}{z} = 1 \, ,
\end{equation}
for $\gamma$ a closed curve encircling the origin, to integrals of differential forms, i.e. 
\begin{equation}\label{eq:naivresidueforms}
\frac{1}{2 \pi i}\int_\gamma \frac{dz\wedge \alpha }{z} =\alpha \,,
\end{equation}
for a smooth differential form $\alpha$. In this way, rational differential forms on $\C^n$ with poles along $z=0$ are naturally identified with smooth forms on the locus $z=0$. The following essentially deals with properly formulating this idea for hypersurfaces $X$ of $\P^{d+1}$. The upshot is that we can write differential forms on $X$ as differential forms with poles on $\P^{d+1}$. 

The setting we will be interested concerns differential forms with poles (`rational forms') on complex projective space. By a result of \cite{griffithsI}, rational $d+1$-forms on $\P^{d+1}$ can always be written in terms of the unique holomorphic $d+1$-form $\Omega_0$ 
\begin{equation}
\varphi = \frac{P(x)}{R(x)} \Omega_0
\end{equation}
for homogeneous polynomials $P(x)$ and $R(x)$ with $\mbox{\deg} R = \mbox{\deg}P + (d+2)$.  
The form $\Omega_0$ is given by
\begin{equation}
\Omega_0 = \sum_{j=0}^{d+1} (-1)^j \,\, x_j \,\,dx_0 \wedge \cdots \wedge \widehat{dx}_j \wedge \cdots \wedge dx_{d+1} \, ,
\end{equation}
where our notation is supposed to indicate that $\widehat{dx}_j$ is omitted from the $\wedge$ product of $dx_j$. There are similar expression for  $n$-forms with $n < d+1$. The constraint on the degrees of $P$ and $R$ hence guarantees that $\varphi$ is invariant under the $\C^*$ acting on the homogeneous coordinates $x_i$. 

Depending on the choice of the denominator, $\varphi$ can have poles of various 
orders along a hypersurface $X \subset \P^{d+1}$. Working modulo exact forms, 
the pole order can sometimes be reduced, as summarized in the following 
statements \cite{griffithsI,griffithsII}.
\begin{itemize}
 \item[a)] For any rational $d+1$-form $\varphi$ there exists a $\eta$ such that $\varphi + d \eta$ has pole order $d+1$. 
 \item[b)] If a rational $d+1$-form $\varphi$ has pole order $k$ along $X$ and there exists and $\eta$ such that $\varphi + d\eta$ has pole order $k-1$, we can choose $\eta$ to have pole order $k-1$. 
 \item[c)] If $X$ is given by $Q(x)=0$ and $\varphi$ has pole order $k$ we can write
 \begin{equation}\label{eq:forms_in_PminusXapp}
  \varphi = \frac{P(x)}{Q(x)^{k}}\,\, \Omega_0 \, .
 \end{equation}
There exists an $\eta$ such that $\varphi + d \eta$ has pole order $k-1$ if and only if $P(x)$ is contained in the Jacobi ideal of $Q(x)$, i.e. the ideal generated by the polynomials $\partial Q / \partial x_j$. 
\end{itemize}

Elements of $H^{d+1}(\P^{d+1}-X)$ can hence be represented by forms such as \eqref{eq:forms_in_PminusXapp}. If $X$ is described by a polynomial $Q(x)=0$ and $Q(x)$ has degree $l$, then $\mbox{deg} P = kl - (d+2)$.  As we always reduce the pole degree of $\varphi$ modulo exact forms, $k$ is at most $d+1$, due to the property a).

The residue map is defined as
\begin{equation}
\mbox{Res} : H^{d+1}(\P^{d+1}-X) \rightarrow H^d_{\mbox{\tiny prim}}(X) 
\end{equation}
as follows. For any $d$-cycle $\Gamma$ on the hypersurface $X$ we set
\begin{equation}
\int_\Gamma \mbox{Res} (\varphi) =  \int_{T(\Gamma)} \varphi\, ,
\end{equation}
where $T(\Gamma)$ is a tube, i.e. a circle bundle over $\Gamma$. It can be shown that such a tube always exists and the definition of the residue is independent of this choice. The image of the residue map is not all of $ H^d(X)$, but only maps to the primitive cohomology $ H^d_{\mbox{\tiny prim}}(X) $, i.e. to those forms perpendicular to the restriction of the hyperplane class. As shown in \cite{griffithsI} the residue map is surjective on the primitive cohomology
\begin{equation}
\mbox{im}\left( \mbox{Res}\right) = H^d_{\mbox{\tiny prim}}(X) \, .
\end{equation}

Let us see in some more detail how this definition of the residue realizes \eqref{eq:naivresidueforms}. Consider a rational $n$-form of pole order $k$ in a small neighborhood containing $Q=0$. There we can choose coordinates such that $\varphi$ becomes
\begin{equation}
\varphi = \frac{dQ \wedge \alpha}{Q^k} + \frac{\beta}{Q^{k-1}} = \frac{1}{k-1}d\left(\frac{\alpha}{Q^{k-1}}\right) + \frac{\beta+\frac{1}{k-1}d\alpha}{Q^{k-1}}
\end{equation}
for some smooth forms $\alpha$ and $\beta$. Hence we may always reduce the pole order of holomorphic forms locally. Using a partition of unity, one can show that this can in fact be done globally, but at the expense of holomorphicity. Iterating this procedure, it follows that we may write (up to exact forms)
\begin{equation}
\varphi = \frac{\gamma \wedge df}{Q} + \delta
\end{equation}
for some smooth forms $\gamma$ and $\delta$. The residue is then simply
\begin{equation}
\mbox{Res}(\varphi) = \gamma|_X \, . 
\end{equation}
This explains why the residue of a holomorphic rational form $\varphi$ such as \eqref{eq:forms_in_PminusXapp} on $\P^n$ is not necessarily holomorphic, except when $k=1$.

Let us now  define $A^{d+1}_k$ to be the additive group of rational $d+1$-forms of pole order at most $k$ along $X$. We can then form the `cohomology groups' 
\begin{equation}
\mathcal{H}_k(X) = \frac{A^{d+1}_k(X)}{d A^{d}_{k-1}(X)} \, .
\end{equation}
The $\mathcal{H}_k(X) $ for different $k$ have a filtration
\begin{equation}
 \mathcal{H}_0 \subset  \mathcal{H}_1 \subset \cdots  \subset  \mathcal{H}_{d+1}\, ,
\end{equation}
which precisely maps to the Hodge filtration of the primitive cohomology under the residue map:
\begin{equation}
\mbox{Res}\left(\mathcal{H}_{k}\right) = \mathcal{F}^{d+1-k} H^{d}_{\mbox{\tiny prim}}(X) \, ,
\end{equation}
where
\begin{equation}
\mathcal{F}^{d+1-k} H^{d} = \bigoplus_{i \geq d+1-k} H^{i,d-i}(X) \, .
\end{equation}
Hence the residue map takes $\mathcal{H}_1(X)$  to $H^{d,0}(X)$, while $\mathcal{H}_2(X)$  maps to $H^{d,0}(X) \oplus H^{d-1,1}(X)$, etc. The forms of maximal pole order, $k=d+1$, are mapped to $\mathcal{F}^{0} H^{d}_{\mbox{\tiny prim}}(X) = H^{d}_{\mbox{\tiny prim}}(X)$.

We can isolate the Hodge cohomology groups of primitive forms by forming the quotients
\begin{equation}\label{eq:hodgevsrationalforms}
H^{p,d-p}_{\mbox{\tiny prim}}(X) = \frac{\mathcal{F}^{p}H^{d}_{\mbox{\tiny prim}}(X)}{\mathcal{F}^{p+1}H^{d}_{\mbox{\tiny prim}}(X)} =  \frac{\mathcal{H}_{d+1-p}(X)}{\mathcal{H}_{d-p}(X)}\, .
\end{equation}
where the last equality is realized by applying the residue map. 

The result of the above is that we can associate Hodge cohomology groups in the middle cohomology with polynomials $P$ of appropriate degree modulo the Jacobi ideal of the polynomial $Q$ defining the hypersurface equation. Consider a rational form of pole degree $k$ written as \eqref{eq:forms_in_PminusXapp}. Such a form defines an element in $A_{k}^{d+1}(X)$ and hence an element in $\mathcal{H}_k(X)$. If $P$ is contained in the Jacobi ideal of $Q$, there exists an $\eta$ such that $\varphi + d \eta$ has a pole of degree $k-1$. This implies that $\varphi$ is equivalent to an element of $A^{d+1}_{k-1}$ in $\mathcal{H}_k(X)$, which in turn implies that $\varphi$ is also contained in $\mathcal{H}_{k-1}(X)$. But this means that $\varphi$ is zero in \eqref{eq:hodgevsrationalforms}, so that the statement we started the paragraph with follows.


\begin{thebibliography}{10}

\bibitem{Becker:1996gj}
K.~Becker and M.~Becker, ``{M theory on eight manifolds},''
  \href{http://dx.doi.org/10.1016/0550-3213(96)00367-7}{{\em Nucl. Phys. B}
  {\bfseries 477} (1996) 155--167},
  \href{http://arxiv.org/abs/hep-th/9605053}{{\ttfamily arXiv:hep-th/9605053}}.

\bibitem{Gukov:1999ya}
S.~Gukov, C.~Vafa, and E.~Witten, ``{CFT's from Calabi-Yau four folds},''
  \href{http://dx.doi.org/10.1016/S0550-3213(00)00373-4}{{\em Nucl. Phys. B}
  {\bfseries 584} (2000) 69--108},
  \href{http://arxiv.org/abs/hep-th/9906070}{{\ttfamily arXiv:hep-th/9906070}}.
  [Erratum: Nucl.Phys.B 608, 477--478 (2001)].

\bibitem{Witten:1996md}
E.~Witten, ``{On flux quantization in M theory and the effective action},''
  \href{http://dx.doi.org/10.1016/S0393-0440(96)00042-3}{{\em J. Geom. Phys.}
  {\bfseries 22} (1997) 1--13},
  \href{http://arxiv.org/abs/hep-th/9609122}{{\ttfamily arXiv:hep-th/9609122}}.

\bibitem{Braun:2008pz}
A.~P. Braun, A.~Hebecker, C.~Ludeling, and R.~Valandro, ``{Fixing D7 Brane
  Positions by F-Theory Fluxes},''
  \href{http://dx.doi.org/10.1016/j.nuclphysb.2009.02.025}{{\em Nucl. Phys. B}
  {\bfseries 815} (2009) 256--287},
  \href{http://arxiv.org/abs/0811.2416}{{\ttfamily arXiv:0811.2416 [hep-th]}}.

\bibitem{Braun:2014xka}
A.~P. Braun and T.~Watari, ``{The Vertical, the Horizontal and the Rest:
  anatomy of the middle cohomology of Calabi-Yau fourfolds and F-theory
  applications},'' \href{http://dx.doi.org/10.1007/JHEP01(2015)047}{{\em JHEP}
  {\bfseries 01} (2015) 047},
\href{http://arxiv.org/abs/1408.6167}{{\ttfamily arXiv:1408.6167 [hep-th]}}.

\bibitem{Braun:2014lwp}
A.~P. Braun and T.~Watari, ``{Distribution of the Number of Generations in Flux
  Compactifications},''
  \href{http://dx.doi.org/10.1103/PhysRevD.90.121901}{{\em Phys. Rev. D}
  {\bfseries 90} no.~12, (2014) 121901},
  \href{http://arxiv.org/abs/1408.6156}{{\ttfamily arXiv:1408.6156 [hep-ph]}}.

\bibitem{Watari:2015ysa}
T.~Watari, ``{Statistics of F-theory flux vacua for particle physics},''
  \href{http://dx.doi.org/10.1007/JHEP11(2015)065}{{\em JHEP} {\bfseries 11}
  (2015) 065}, \href{http://arxiv.org/abs/1506.08433}{{\ttfamily
  arXiv:1506.08433 [hep-th]}}.

\bibitem{Braun:2014ola}
A.~P. Braun, Y.~Kimura, and T.~Watari, ``{The Noether-Lefschetz problem and
  gauge-group-resolved landscapes: F-theory on K3 $\times$ K3 as a test
  case},'' \href{http://dx.doi.org/10.1007/JHEP04(2014)050}{{\em JHEP}
  {\bfseries 04} (2014) 050}, \href{http://arxiv.org/abs/1401.5908}{{\ttfamily
  arXiv:1401.5908 [hep-th]}}.

\bibitem{Bizet:2014uua}
N.~Cabo~Bizet, A.~Klemm, and D.~Vieira~Lopes, ``{Landscaping with fluxes and
  the E8 Yukawa Point in F-theory},''
  \href{http://arxiv.org/abs/1404.7645}{{\ttfamily arXiv:1404.7645 [hep-th]}}.

\bibitem{Aspinwall:2005ad}
P.~S. Aspinwall and R.~Kallosh, ``{Fixing all moduli for M-theory on K3xK3},''
  \href{http://dx.doi.org/10.1088/1126-6708/2005/10/001}{{\em JHEP} {\bfseries
  10} (2005) 001}, \href{http://arxiv.org/abs/hep-th/0506014}{{\ttfamily
  arXiv:hep-th/0506014}}.

\bibitem{Braun:2011zm}
A.~P. Braun, A.~Collinucci, and R.~Valandro, ``{G-flux in F-theory and
  algebraic cycles},''
  \href{http://dx.doi.org/10.1016/j.nuclphysb.2011.10.034}{{\em Nucl. Phys.}
  {\bfseries B856} (2012) 129--179},
\href{http://arxiv.org/abs/1107.5337}{{\ttfamily arXiv:1107.5337 [hep-th]}}.

\bibitem{movasati_loyola_17}
H.~{Movasati} and R.~{Villaflor Loyola}, ``{Periods of linear algebraic
  cycles},'' {\em arXiv e-prints} (Apr., 2017) arXiv:1705.00084,
  \href{http://arxiv.org/abs/1705.00084}{{\ttfamily arXiv:1705.00084
  [math.AG]}}.

\bibitem{2018arXiv181203964V}
R.~{Villaflor Loyola}, ``{Periods of Complete Intersection Algebraic Cycles},''
  {\em arXiv e-prints} (Dec., 2018) arXiv:1812.03964,
  \href{http://arxiv.org/abs/1812.03964}{{\ttfamily arXiv:1812.03964
  [math.AG]}}.

\bibitem{Haack:2001jz}
M.~Haack and J.~Louis, ``{M theory compactified on Calabi-Yau fourfolds with
  background flux},''
  \href{http://dx.doi.org/10.1016/S0370-2693(01)00464-6}{{\em Phys. Lett. B}
  {\bfseries 507} (2001) 296--304},
  \href{http://arxiv.org/abs/hep-th/0103068}{{\ttfamily arXiv:hep-th/0103068}}.

\bibitem{Giddings:2001yu}
S.~B. Giddings, S.~Kachru, and J.~Polchinski, ``{Hierarchies from fluxes in
  string compactifications},''
  \href{http://dx.doi.org/10.1103/PhysRevD.66.106006}{{\em Phys. Rev. D}
  {\bfseries 66} (2002) 106006},
  \href{http://arxiv.org/abs/hep-th/0105097}{{\ttfamily arXiv:hep-th/0105097}}.

\bibitem{Denef:2008wq}
F.~Denef, ``{Les Houches Lectures on Constructing String Vacua},'' {\em Les
  Houches} {\bfseries 87} (2008) 483--610,
  \href{http://arxiv.org/abs/0803.1194}{{\ttfamily arXiv:0803.1194 [hep-th]}}.

\bibitem{Strominger:1990pd}
A.~Strominger, ``{SPECIAL GEOMETRY},''
  \href{http://dx.doi.org/10.1007/BF02096559}{{\em Commun. Math. Phys.}
  {\bfseries 133} (1990) 163--180}.

\bibitem{Greene:1993vm}
B.~R. Greene, D.~R. Morrison, and M.~R. Plesser, ``{Mirror manifolds in higher
  dimension},'' \href{http://dx.doi.org/10.1007/BF02101657}{{\em Commun. Math.
  Phys.} {\bfseries 173} (1995) 559--598},
  \href{http://arxiv.org/abs/hep-th/9402119}{{\ttfamily arXiv:hep-th/9402119
  [hep-th]}}.
[AMS/IP Stud. Adv. Math.1,745(1996)].

\bibitem{voisin2010hodge}
C.~Voisin, ``Hodge loci,'' {\em Handbook of moduli} {\bfseries 3} (2010)
  507--546.

\bibitem{movasati_book}
H.~Movasati, {\em Hodge Theory}.
\newblock International Press, Boston, 2020.
\newblock \url{http://w3.impa.br/%7Ehossein/myarticles/hodgetheory.pdf}.

\bibitem{Aoki1983}
N.~Aoki and T.~Shioda, {\em Generators of the N{\'e}ron-Severi Group of a
  Fermat Surface},
  \href{http://dx.doi.org/10.1007/978-1-4757-9284-3_1}{pp.~1--12}.
\newblock Birkh{\"a}user Boston, Boston, MA, 1983.
\newblock \url{https://doi.org/10.1007/978-1-4757-9284-3_1}.

\bibitem{shioda1982picard}
T.~Shioda, ``On the picard number of a fermat surface,'' {\em J. Fac. Sci.
  Univ. Tokyo} {\bfseries 28} (1982) 725--734.

\bibitem{aoki1987_alg_cycles}
N.~AOKI, ``Some new algebraic cycles on fermat varieties,''
  \href{http://dx.doi.org/10.2969/jmsj/03930385}{{\em J. Math. Soc. Japan}
  {\bfseries 39} no.~3, (07, 1987) 385--396}.
  \url{https://doi.org/10.2969/jmsj/03930385}.

\bibitem{shioda_hodge_conj_fermat_79}
T.~Shioda, ``The hodge conjecture for fermat varieties,''
  \href{http://dx.doi.org/10.1007/BF01428804}{{\em Mathematische Annalen}
  {\bfseries 245} (11, 1979) 175--184}.

\bibitem{ASENS_1969_4_2_4_583_0}
N.~Katz, ``On the intersection matrix of a hypersurface,''
  \href{http://dx.doi.org/10.24033/asens.1185}{{\em Annales scientifiques de
  l'\'Ecole Normale Sup\'erieure} {\bfseries Ser. 4, 2} no.~4, (1969)
  583--598}.

\bibitem{ogus_crystal}
A.~Ogus, ``Griffiths transversality in crystalline cohomology,'' {\em Annals of
  Mathematics} {\bfseries 108} no.~3, (1978) 395--419.
  \url{http://www.jstor.org/stable/1971182}.

\bibitem{Ran1980}
Z.~Ran, ``Cycles on fermat hypersurfaces,'' {\em Compositio Mathematica}
  {\bfseries 42} no.~1, (1980) 121--142.

\bibitem{griffiths_carlson_83}
J.~Carlson, M.~Green, P.~A. Griffiths, and J.~Harris, ``Infinitesimal
  variations of hodge structure (i)),'' {\em Compositio Mathematica} {\bfseries
  50} no.~2-3, (1983) 109--205.

\bibitem{voisin_2003}
C.~Voisin, \href{http://dx.doi.org/10.1017/CBO9780511615177}{{\em Hodge Theory
  and Complex Algebraic Geometry II}}, vol.~2 of {\em Cambridge Studies in
  Advanced Mathematics}.
\newblock Cambridge University Press, 2003.

\bibitem{2014arXiv1411.1766M}
H.~{Movasati}, ``{Gauss-Manin connection in disguise: Noether-Lefschetz and
  Hodge loci},'' {\em arXiv e-prints} (Nov., 2014) arXiv:1411.1766,
  \href{http://arxiv.org/abs/1411.1766}{{\ttfamily arXiv:1411.1766 [math.AG]}}.

\bibitem{Greene:1990ud}
B.~R. Greene and M.~Plesser, ``{Duality in {Calabi-Yau} Moduli Space},''
  \href{http://dx.doi.org/10.1016/0550-3213(90)90622-K}{{\em Nucl. Phys. B}
  {\bfseries 338} (1990) 15--37}.

\bibitem{2016arXiv160206607M}
H.~{Movasati}, ``{Why should one compute periods of algebraic cycles?},'' {\em
  arXiv e-prints} (Feb., 2016) arXiv:1602.06607,
  \href{http://arxiv.org/abs/1602.06607}{{\ttfamily arXiv:1602.06607
  [math.AG]}}.

\bibitem{Giryavets:2003vd}
A.~Giryavets, S.~Kachru, P.~K. Tripathy, and S.~P. Trivedi, ``{Flux
  compactifications on Calabi-Yau threefolds},''
  \href{http://dx.doi.org/10.1088/1126-6708/2004/04/003}{{\em JHEP} {\bfseries
  04} (2004) 003}, \href{http://arxiv.org/abs/hep-th/0312104}{{\ttfamily
  arXiv:hep-th/0312104}}.

\bibitem{Denef:2004dm}
F.~Denef, M.~R. Douglas, and B.~Florea, ``{Building a better racetrack},''
  \href{http://dx.doi.org/10.1088/1126-6708/2004/06/034}{{\em JHEP} {\bfseries
  06} (2004) 034}, \href{http://arxiv.org/abs/hep-th/0404257}{{\ttfamily
  arXiv:hep-th/0404257}}.

\bibitem{Louis:2012nb}
J.~Louis, M.~Rummel, R.~Valandro, and A.~Westphal, ``{Building an explicit de
  Sitter},'' \href{http://dx.doi.org/10.1007/JHEP10(2012)163}{{\em JHEP}
  {\bfseries 10} (2012) 163}, \href{http://arxiv.org/abs/1208.3208}{{\ttfamily
  arXiv:1208.3208 [hep-th]}}.

\bibitem{Cicoli:2013cha}
M.~Cicoli, D.~Klevers, S.~Krippendorf, C.~Mayrhofer, F.~Quevedo, and
  R.~Valandro, ``{Explicit de Sitter Flux Vacua for Global String Models with
  Chiral Matter},'' \href{http://dx.doi.org/10.1007/JHEP05(2014)001}{{\em JHEP}
  {\bfseries 05} (2014) 001}, \href{http://arxiv.org/abs/1312.0014}{{\ttfamily
  arXiv:1312.0014 [hep-th]}}.

\bibitem{ALJOVIN2019177}
E.~Aljovin, H.~Movasati, and R.~{Villaflor Loyola}, ``Integral hodge conjecture
  for fermat varieties,''
  \href{http://dx.doi.org/https://doi.org/10.1016/j.jsc.2019.02.006}{{\em
  Journal of Symbolic Computation} {\bfseries 95} (2019) 177 -- 184},
  \href{http://arxiv.org/abs/1711.02628}{{\ttfamily arXiv:1711.02628
  [math.AG]}}.

\bibitem{Schimmrigk:2008mp}
R.~Schimmrigk, ``{Emergent spacetime from modular motives},''
  \href{http://dx.doi.org/10.1007/s00220-010-1179-4}{{\em Commun. Math. Phys.}
  {\bfseries 303} (2011) 1--30},
  \href{http://arxiv.org/abs/0812.4450}{{\ttfamily arXiv:0812.4450 [hep-th]}}.

\bibitem{Kachru:2020sio}
S.~Kachru, R.~Nally, and W.~Yang, ``{Supersymmetric Flux Compactifications and
  Calabi-Yau Modularity},'' \href{http://arxiv.org/abs/2001.06022}{{\ttfamily
  arXiv:2001.06022 [hep-th]}}.

\bibitem{Schimmrigk:2020dfl}
R.~Schimmrigk, ``{Flux vacua and modularity},''
  \href{http://arxiv.org/abs/2003.01056}{{\ttfamily arXiv:2003.01056
  [hep-th]}}.

\bibitem{livne_K3}
R.~Livn\'e, ``{Motivic orthogonal two-dimensional representations of
  $\mbox{Gal}\,(\bar{\mathbb{Q}}/\mathbb{Q})$},'' {\em Israel J. Math.} no.~92,
  (1995) 149--156. \url{https://doi.org/10.1007/BF02762074}.

\bibitem{Candelas:2019llw}
P.~Candelas, X.~de~la Ossa, M.~Elmi, and D.~Van~Straten, ``{A One Parameter
  Family of Calabi-Yau Manifolds with Attractor Points of Rank Two},''
  \href{http://arxiv.org/abs/1912.06146}{{\ttfamily arXiv:1912.06146
  [hep-th]}}.

\bibitem{griffithsI}
P.~A. Griffiths, ``On the periods of certain rational integrals: I,'' {\em
  Annals of Mathematics} {\bfseries 90} no.~3, (1969) 460--495.
  \url{http://www.jstor.org/stable/1970746}.

\bibitem{griffithsII}
P.~A. Griffiths, ``On the periods of certain rational integrals: Ii,'' {\em
  Annals of Mathematics} {\bfseries 90} no.~3, (1969) 496--541.
  \url{http://www.jstor.org/stable/1970747}.

\bibitem{cox1999mirror}
D.~Cox and S.~Katz, {\em Mirror Symmetry and Algebraic Geometry}.
\newblock Mathematical surveys and monographs. American Mathematical Society,
  1999.

\bibitem{Candelas:2000fq}
P.~Candelas, X.~de~la Ossa, and F.~Rodriguez-Villegas, ``{Calabi-Yau manifolds
  over finite fields. 1.},''
  \href{http://arxiv.org/abs/hep-th/0012233}{{\ttfamily arXiv:hep-th/0012233}}.

\bibitem{Doran:2007jw}
C.~Doran, B.~Greene, and S.~Judes, ``{Families of quintic Calabi-Yau 3-folds
  with discrete symmetries},''
  \href{http://dx.doi.org/10.1007/s00220-008-0473-x}{{\em Commun. Math. Phys.}
  {\bfseries 280} (2008) 675--725},
  \href{http://arxiv.org/abs/hep-th/0701206}{{\ttfamily arXiv:hep-th/0701206}}.

\end{thebibliography}

\providecommand{\href}[2]{#2}\begingroup\raggedright\endgroup

\end{document}